\documentclass{amsart}
\usepackage{graphicx}
\usepackage{color}
\usepackage{tikz}
\usepackage{amssymb}
\usepackage{amsfonts}
\usepackage{amsmath}
\usepackage{amsthm}
\newtheorem{Theorem}{Theorem}[part]
\newtheorem{Definition}{Definition}[part]
\newtheorem{Proposition}{Proposition}[part]

\newtheorem{Lemma}{Lemma}[part]
\newtheorem{Corollary}{Corollary}[part]
\newtheorem{Remark}{Remark}[part]
\newtheorem{Example}{Example}[part]

\newcommand{\be}{\begin{equation}}
\newcommand{\ee}{\end{equation}}

\numberwithin{equation}{section}

\begin{document}

\title{Market selection with learning and catching up with the Joneses}
\thanks{This paper was previously circulated under the title "Natural selection with habits and learning in heterogeneous economies".}
 \author{Roman Muraviev \\
 Department of Mathematics and RiskLab\\
 ETH Zurich\\
 }
\date{January 9, 2012.}
\address{
 Department of Mathematics and RiskLab, ETH, Zurich 8092, Switzerland
 \newline
 {e.mail: roman.muraviev@math.ethz.ch}}

\begin{abstract}
We study the market selection hypothesis in complete financial markets, populated by heterogeneous agents.
We allow for a rich structure of heterogeneity: individuals may differ in their beliefs concerning the economy,
information and learning mechanism, risk aversion, impatience and 'catching up with Joneses'
preferences. We develop new techniques for studying the long-run behavior of such economies, based on the Strassen's
functional law of iterated logarithm. In particular, we explicitly determine an agent's survival index and show how
the latter depends on the agent's characteristics. We use these results to study the long-run behavior of the
equilibrium interest rate and the market price of risk.
\end{abstract}

\subjclass[2010]{Primary: 91B69 Secondary: 91B16.}
\keywords{Natural Selection, Heterogeneous Equilibrium, Diverse Beliefs, Learning, 
Survival Index, Catching up with The Joneses} \maketitle
\markboth{R. Muraviev}{Market selection with learning and catching up with the Joneses}

\renewcommand{\theequation}{\arabic{section}.\arabic{equation}}
\pagenumbering{arabic}

\section{Introduction}

A fundamental question in the modern theory of financial economics is concerned with the so-called market selection hypothesis,
dating back to the ideas of Friedman \cite{Fr}. Motivated by the postulate that agents with inaccurate forecasts will eventually
be driven out of the economy, this hypothesis can be stated informally as "If you are so smart, why aren't you rich?". Formally,
market selection in financial markets examines the agents' long-run survival\footnote{An agent is said to survive in the long-run if
the ratio of his consumption to the aggregate consumption stays positive with positive probability as time goes to infinity.} capability
and price impact in equilibrium models. There is a vast body of literature dealing with this topic; see e.g. Blume and Easley \cite{BE},
Cvitani\'c, Jouini, Malamud and Napp \cite{CJMN}, Nishide and Rogers \cite{NR}, Sandroni \cite{Sa}, and Yan \cite{Y1,Y2}.

This paper investigates the market selection hypothesis (or, natural selection, for short) and the long-run behavior of asset prices in a
complete market setting with highly heterogeneous investors. Individuals may differ in their beliefs concerning the economy,
information and learning mechanism, risk aversion, impatience (time preference rate) and degree of habit-formation.
Each individual in our model is represented by a generalized version of the \textit{catching up with the Joneses}
power utility function of Chan and Kogan \cite{CK}. This model of preferences is sometimes referred to in the literature as
\textit{exogenous habit-formation}, since it incorporates the impact of a certain given stochastic process on the individual's
consumption policy. Agents are assumed to possess only partial information regarding the events associated with the evolution
of the market. More precisely, the stochastic dynamics of the mean growth-rate of the economy\footnote{We assume that the mean
growth-rate follows an Ornstein-Uhlenbeck process.} are \textit{unobservable}, and the agents' information set consists of the
aggregate endowment and a publicly observable signal. Furthermore, agents are allowed to have \textit{diverse beliefs} concerning
the values of the initial and average mean growth-rate. Individuals may be irrational in the way they interpret the public signal:
some of them may be over- (or, under-)confident about the informativeness of the public signal. We use the standard way of
modeling over-(or, under-)confidence, originated in Dumas, Kurshev, and Uppal \cite{DKU} and Scheinkman and Xiong \cite{SX}:
we assume that agents' beliefs concerning the instantaneous correlation of the public signal with the economy's growth-rate may
differ from its actual value.\footnote{This is a realistic assumption as correlations are extremely difficult to estimate empirically.}
The agents are rational in the sense that they use a standard Kalman filter to update their expectations about the economy's growth-rate.
The heterogeneous filtering rules yield highly non-trivial dynamics for the individual consumption and the equilibrium state price density,
determined by the market clearing condition. In particular, subjective probability densities describing the agents' beliefs give rise to multiple
new state variables, which govern the dynamics of the economy. We refer to Back \cite{Back} for a survey on filtering and incomplete
information in asset pricing theory.

Let us describe the contribution of this work to the literature on equilibrium and natural selection.
Firstly, as described above, we analyze a very general paradigm of heterogeneous economies including diverse beliefs,
Kalman filtering and exogenous state-dependent habit formation preferences. We provide a comprehensive description
of the equilibrium characteristics, that can be used for further research in other possible directions.
Secondly, this complex setting in turn allows detecting which traits (both behavioral-preferential and information-related)
are beneficial for survival. That is, as in Yan \cite{Y1}, we reveal that there is a unique surviving agent in the long-run. Moreover,
we show that the interest rate and the market price of risk behave asymptotically as those of an economy, populated solely by this surviving agent.
Lastly, to derive our results, we develop new techniques based mainly on the Strassen's functional law of iterated logarithm.
To the best of our knowledge, these methods have never been used in the general equilibrium literature before.

The conclusions and implications on natural selection are as follows. Most importantly, our findings indeed confirm, to a large extent, the validity of
the market selection hypothesis.

In a growing economy, the less effectively risk-averse\footnote{In our model, the effective risk aversion depends on the level
of habit-formation (see (\ref{effective_ra}))} agent is the one to survive in the long-run. This result is consistent with previous studies
(see e.g. Cvitani\'c, Jouini, Malamud, and Napp \cite{CJMN}). However, the impact of habit-formation on the effective risk aversion,
and thus in particular on survival, is quite novel. As it turns out, if the (standard non-effective) risk aversion coefficient is
above one, then the individual with the strongest habit will survive. Intuitively, this makes sense, as aggressiveness in a growing
economy among somewhat moderate individuals is supposed to be a plus. On the other hand, if the (standard non-effective) level of risk aversion
is below one (i.e., individuals are relatively risk-seeking in the classical sense), the agent with the lowest degree of
habit-formation will dominate. This is not surprising at all, as excess aggressiveness can cause bubbles leading to extinction.

Some of our conclusions concerning the interaction of diverse beliefs and survival are quite intriguing,
and seem to be quite hard to predict without a delicate analysis.

When agents differ only in their beliefs concerning the average mean growth-rate,
the one with the most accurate forecast will dominate the market, as expected.
If all agents are over-confident (or under-confident), then, again, the agent with the
best guess will beat the others. However, if some agents are over-confident and others are
under-confident, the situation is more complex.
For instance, it may happen that in a situation where the public signal provides some relevant
information about the market,
the surviving agent will be the one who (wrongly) believes that this signal is a pure noise, whereas the agent who is
significantly overconfident in the informativeness of the signal will be eliminated from the economy.
Furthermore, in some cases, agents that believe in a negative correlation of the signal will survive, while
individuals who believe in a (too high) positive correlation will be extinct, despite an actual positive correlation.
See Figure \ref{fig:surv} for an example describing these phenomena.
Even though it is somewhat debatable which property of the preceding two can be considered a more rational one,
we still learn that theoretically, the market selection hypothesis is valid, at least in some modified form.

We now review some related works. The most closely related to ours are the papers by Yan \cite{Y1} and Cvitani\'c, Jouini,
Malamud and Napp  \cite{CJMN}.\footnote{Bhamra and Uppal \cite{BU}, Dumas \cite{D}, and Wang \cite{W} considered the same model,
but with only two agent types and heterogeneity coming only from risk aversion.} Specifically, these authors consider a special case of
our model corresponding to the case when there is no learning and agents having standard CRRA preferences without any habit formation.
In terms of modeling heterogeneous beliefs and learning, our model closely follows the one of Dumas, Kurshev, and Uppal \cite{DKU}
and Scheinkman and Xiong \cite{SX}, who considered a special case of our model: a two-agent economy with standard CRRA utility functions,
and the public signal being a pure noise, uncorrelated with the economy's growth-rate. Chan and Kogan \cite{CK} consider a special case of
our model with homogeneous 'catching up with the Joneses' habit levels and a continuum of agents with heterogeneous risk aversions.
Xiouros and Zapatero \cite{XZ} derive a closed form expression for the equilibrium state price density in the Chan and Kogan \cite{CK}
model. Cvitani\'c and Malamud \cite{CM} study how long-run risk sharing depends on the presence of multiple agents with different levels
of risk aversion. Kogan, Ross, Wang and Westerfield \cite{KRWW1} and Cvitani\'c and Malamud \cite{CM2} study the interaction of survival
and price impact in economies where agents derive utility only from terminal consumption. Fedyk, Heyerdahl-Larsen and Walden \cite{FHLW}
extend the model of Yan \cite{Y1} by allowing for many assets. Kogan, Ross, Wang and Westerfield \cite{KRWW2} study the link between survival
and price impact in the presence of intermediate consumption, and allow for general utilities with unbounded relative risk aversion and a general
dividend process. Another quite significant direction of the complete market risk sharing literature concentrates on the equilibrium effects of
heterogeneous beliefs. Bhamra and Uppal \cite{BU2} derive a characterization of the equilibrium state price density by means of infinite
series that admits
a closed form solution for specific coefficients, in a two-agent economy with diverse beliefs and heterogenous CRRA preferences.
With CRRA agents differing only in their beliefs, the equilibrium state price density can be derived in a closed form, and thus many
equilibrium properties can be analyzed in detail. See, e.g., Basak \cite{Basak1,Basak2}, Jouini and Napp
\cite{JN1,JN2}, Jouini, Martin and Napp \cite{JMN} and Xiong and Yan \cite{XY}.

The paper is organized as follows. In Section 2 we introduce the model and provide some
preliminary results. Section 3 is devoted to
a brief description of the equilibrium state price density in homogeneous and heterogeneous
settings. In Section 4, we present the main result of the paper and discuss some implications.
Section 5 deals with some auxiliary results that are crucial for the proof of the main result.
In Section 6 we prove the main result. Finally, in Section 7 we establish long-run results for
the interest rate and the market price of risk. Some of the results appearing in sections 5 and 7 are of an independent mathematical interest.

\section{Preliminaries }
We consider a continuous-time Arrow-Debreu economy with an
infinite horizon, in which heterogeneous agents maximize their
utility functions from consumption. The uncertainty in our model
is captured by a (complete) probability space
$ \left( \Omega, \mathcal{F}_{ \infty } , P \right)  $ and a continuous  filtration
 $ \mathcal{F} :=  \left( \mathcal{F}_t \right)_{ t \in [ 0 , \infty)  } $, with $ \mathcal{F }_0 = \{ \emptyset , \Omega \} $. We fix three standard and independent Wiener processes
$ ( W^{(i)}_{ t })_{ t \in [0 , \infty)  } $, $i=1,2,3$, adapted to the filtration $\mathcal{F}$.
There are $N$ different types of agents in the economy, labeled by $i=1,...,N.$
Each agent $i$ is equipped with a non-negative endowment
process $ \left( \epsilon^{i}_t \right)_{ t \in [0 , \infty) } $ adapted to the filtration $\mathcal{G}$ (see (\ref{G_filtration})). We denote by
$ D_t := \sum_{i=1 }^{N} \epsilon^{i}_t $ the \textit{aggregate endowment} process and assume that $ (
D_t )_{ t \in [ 0 , \infty )  } $ satisfies
\begin{equation}
\frac{d D_t }{ D_t  } =  \mu^{D}_t dt + \sigma^{D} dW^{(1)}_t \ , \ D_0 =1,
\label{dividend}
\end{equation}
or equivalently
\begin{equation}
D_t = \exp \left(  \int_{0}^{t} \mu^{D}_s ds - \frac{1}{2} ( \sigma^{D} )^2 t   + \sigma^{D} W^{(1)}_t   \right),
\label{divident_explicitly}
\end{equation}
where the constant $\sigma^{D}>0$ represents the volatility.
The \textit{mean growth-rate} $ ( \mu^{D}_t )_{ t \in [  0 , \infty) } $ is an Ornstein-Uhlenbeck process
that solves uniquely the SDE
\begin{equation}
d \mu^{D}_t = - \xi ( \mu^{D}_t - \overline{ \mu} ) dt + \sigma^{ \mu} d W^{(2)}_t  \ , \ \mu^{D}_0 = \mu,
\label{mgr}
\end{equation}
that is
\begin{equation}
\mu^{D}_t =  \overline{ \mu }  + \left( \mu_0 -  \overline{ \mu }   \right)  e^{ - \xi t } + \sigma^{ \mu}   e^{ - \xi t   } \int_{0}^{t} e^{ \xi  s  } d W^{(2)}_{s},
\label{mgr_explicit}
\end{equation}
where $ \overline{ \mu} , \mu_0 $ and $ \sigma^{ \mu} $ are some real numbers and $ \xi > 0  $.
The numbers $\overline{ \mu} , \mu_0$ will be referred to as the \textit{average} and \textit{initial} mean growth-rate, respectively.

\subsection{The Financial Market}
We consider a financial market that consists of at least two long-lived
securities: a risky stock $ ( S_t)_{
t \in [ 0 , \infty)  } $ and a bank account $ ( S_t^0)_{
t \in [ 0 , \infty)  } $. In addition to this, there are other (not explicitly modeled) assets
guaranteeing that the market is dynamically complete\footnote{
In this setting, the model can be implemented by a complete
securities market with a unique state price density derived in
equilibrium (as for instance in Duffie and Huang \cite{DH}). More specifically, the filtration
$\mathcal{G}$ is generated by the Brownian motion $s_t$ (which is interpreted as a public signal)
and the aggregate endowment process $D_t$. Nevertheless, as explained in Remark \ref{filtration_rem},
the filtration $\mathcal{G}$ is also generated by the Brownian motions $s_t$ and $ W^{(0)}_t $.
Thereby, the market can be completed by adding one additional security to $ S_t$.
However, since the price of this security would be determined endogenously, one would have to verify endogenous completeness.
This can be done by using the techniques of Hugonnier, Malamud and Trubowitz \cite{HMT}. Otherwise, we can just assume that
there are sufficiently many (derivative) assets, completing the market.}
for $\mathcal{G}$ adapted claims (where the filtration $\mathcal{G} := \left( \mathcal{G}_t \right)_{t \in [0, \infty)} $ is defined in (\ref{G_filtration})).
We emphasize that this filtration coincides with the symmetric information shared by all agents.
The bond is in zero net supply and the stock is a claim to the total endowment of
the economy $ ( D_t )_{ t \in [ 0 , \infty )  } $ and has a net
supply of one share. The risk-less bond is given by $S^{0}_{t} = e^{
\int_{0}^{t} r_s  ds } $, where $( r_t )_{ t \in [ 0 , \infty) } $ is
the risk-free rate process. We assume a \textit{unique positive state price density} denoted by $( M_t )_{t
\in [ 0 , \infty) } ,$ that is, a positive adapted process to $\mathcal{G}$ that satisfies
\[
M_{t}  = E \left[ e^{ \int_{ t}^{ u} r_s ds } M_u   \big| \mathcal{G}_t  \right] ,
\]
for all $u > t $, and
\[
S_{t}  = E \left[ \int_{t}^{ \infty} \frac{ M_u }{ M_t} D_u du     \big| \mathcal{G}_t  \right],
\]
for all $t > 0 $. Note that our assumption excludes arbitrage opportunities in the model.
The state price density, as well as all other parameters, are to be derived
endogenously in equilibrium.

\subsection{Preferences and Equilibrium}
Agent $ i $ is maximizing his intertemporal von Neumann-Morgenstern expected utility
\[
\sup_{ ( c_{it} )_{t \in [ 0 , \infty)  }  } E^{P^{i}} \left[ \int_{0}^{ \infty } e^{ - \rho_i t} U_{i} (
c_{it} ) dt \right],
\]
from consumption, under the constraints that the consumption stream $ ( c_{it} )_{ t \in [ 0 , \infty  ) } $ is a positive process adapted to $ \mathcal{G}$ (which is defined in (\ref{G_filtration}))
and lies in the budget set
\[
E \left[ \int_{ 0}^{ \infty } c_{it} M_t  dt \right] \leq E \left[ \int_{
0}^{ \infty } \epsilon^{i}_t M_t dt \right] .
\]
Here, $ E^{ P_i} [ \cdot ]  $ stands for the expectation with
respect to the subjective probability measure $ P_i $ of agent $i$.
The exact form of $P_i$ is specified in $(\ref{A})$. We assume
that all agents are represented by 'catching up with the Joneses\footnote{This paradigm of a utility function was first introduced in Abel \cite{Abel}, and is commonly referred to in the literature as a utility with exogenous habits. This specification describes a decision maker who experiences an impact of the 'standard of living' index.}' preferences:
\[
U_{i} \left( c_{it} \right) = \frac{1}{1 - \gamma_i } \left( \frac{ c_{it} }{ H_{
i t } } \right)^{ 1 - \gamma_i }.
\]
The subjective 'standard of living' index $ \left( H_{it} \right)_{ t \in [ 0 , \infty) }$ is defined through a certain geometric
average of the aggregate endowment process.
We consider here a more general specification for
$H_{it}$ than the one in Chan and Kogan \cite{CK}.
Namely, we set $ H_{it} = e^{ \beta_i
 x_{t} }, $ for some $ \beta_i \geq 0$, where
\begin{equation}
x_{t} = e^{ - \lambda t} \left( x_{0}  + \lambda
\int_{0}^{t} e^{ \lambda s  } \log(D_s) ds \right),
\label{standard_living}
\end{equation}
or equivalently, $( x_t)_{ t \in [ 0 , \infty) } $ solves the SDE
\[
d x_t = \lambda ( \log(D_t) - x_t  ) dt.
\]
For each agent $i$, the number $ \beta_i $ measures the impact of the index $x_t$ on the agent; in particular,
when $ \beta_i = 0 $, the agent is not influenced by the index at all. For large $ \beta_i $, the influence is somewhat heavy. In complete markets, the optimal consumption stream can be easily derived as in the following statement.

\begin{Proposition} The optimal consumption stream of agent
$i$, in a complete market represented by a state price density $(M_t )_{t \in [ 0 , \infty)},$
is given by
\[
c_{it} = e^{ \frac{ \rho_i }{ \gamma_i } t }
M^{ - \frac{1 }{ \gamma_i } }_{t} Z^{ \frac{1}{ \gamma_i }}_{it}
H^{ \frac{ \gamma_i - 1 }{ \gamma_i } }_{it}  c_{i0} ,
\]
and
\[
E \left[ \int_{ 0}^{ \infty } c_{it} \xi_t \right] = E \left[ \int_{
0}^{ \infty } \epsilon^{i}_t M_t \right],
\]
where the density process $ ( Z_{it} )_{ t \in [0, \infty)} $ is given in (\ref{A}).
\label{consumption_explicit}
\end{Proposition}
\textbf{Proof.} The assertion follows by standard duality arguments involving the first-order conditions. $ \qed$
\newline
\newline
Finally, we introduce the notion of Arrow-Debreu equilibrium.

\begin{Definition} An \textit{equilibrium} is a pair $ \left( (c_{it})_{ t \in [ 0 , \infty) } , (M_t)_{ t \in [ 0 , \infty) } \right) $ such that:
\newline
a. Each process $(c_{it})_{t \in [0 , \infty) } $ is the
optimal consumption stream of agent $i$ and $(M_t)_{ t \in [ 0 , \infty) } $ is the
state price density that represents the market.
\newline
b. The market clearing condition is satisfied:
\begin{equation}
\sum_{i=1}^{N} c_{it} = D_t,
\label{MarketC}
\end{equation}
for all $t > 0$.
\end{Definition}

\subsection{Diverse Beliefs and Learning }
The are two processes in the economy that are \textit{observable} by all agents. The first one is the aggregate endowment process $(D_{t})_{t \in [0,\infty)}$, and the second one is a certain public signal
\[
s_t = \phi W^{(2)}_t + \sqrt{ 1 - \phi^2 } W^{(3)}_t ,
\]
for some $ \phi \in [0,1).$ That is, the public signal exhibits a
non-negative correlation  $\phi \in [0,1)$ with the shock governing the mean growth-rate process.
The corresponding filtration is denoted by
\begin{equation}
\mathcal{G}_t := \sigma \left( \left\{ s_u  ; u \leq t \right\} \bigcup \left\{
D_{u} ; u \leq t  \right\} \right).
\label{G_filtration}
\end{equation}
In contrast to this, the mean growth-rate process is \textit{unobservable}. That is,
neither of the agents possesses access to the
data revealing the dynamics of the process $ (
\mu^{D}_t)_{ t \in [ 0 , \infty) } $. Furthermore, agents may have
diverse beliefs concerning the average and initial mean growth-rate.
More precisely, each agent $i$ believes that the initial
mean growth-rate is some $  \mu_{0i} \in \mathbb{R} $ and that the
average mean growth-rate is some $ \overline{\mu}_{i} \in
\mathbb{R} $. That is to say, before filtering, agent $i$ assigns in his mind the
following model for $ \mu^{D}_t$:
\begin{equation}
\overline{ \mu }_i  + \left( \mu_{i0} -  \overline{ \mu }_i   \right)  e^{ - \xi t } +
\sigma^{ \mu}   e^{ - \xi t   } \int_{0}^{t} e^{ \xi  s  } d W^{(2)}_{s}.
\label{change1}
\end{equation}
Furthermore, individuals may have an irrational perception of the signal.
Concretely, each agent $i$ believes that the public signal $( s_t )_{t \in [0, \infty)} $
has a correlation $\phi_{i} \in [-1,1)$ with $ ( W^{(2)}_t )_{t \in [0 , \infty) }$, when if fact,
the correlation is $\phi \in [0,1)$. Therefore,
under the belief of agent $i$, the following model is attributed to the signal $s_t$:
\begin{equation}
\phi_i W^{(2)}_t + \sqrt{ 1 - \phi^2_i } W^{(3)}_t .
\label{signal_belief}
\end{equation}
We denote by $Q^{i}$ the measure corresponding to agent's $i-$th beliefs regarding the models in (\ref{change1}) and (\ref{signal_belief}), where $W^{(1)}_t, W^{(2)}_t$ and $W^{(3)}_t$ are independent Wiener processes under $Q^{i}$. Consequently, agents are in the process of learning and filtering out the dynamics of the mean growth-rate, which is deduced by using the
theory of optimal filtering.

\begin{Definition}
The process
\[
\mu^{D}_{it} := E^{Q^{i}} \left[  \overline{ \mu }_i  + \left( \mu_{i0} -  \overline{ \mu }_i   \right)  e^{ - \xi t } +
\sigma^{ \mu}   e^{ - \xi t   } \int_{0}^{t} e^{ \xi  s  } d W^{(2)}_{s}  \big| \mathcal{G}_t \right]
\]
is called the \textit{subjective mean growth-rate} of agent $i$.
\label{def1}
\end{Definition}

\begin{Proposition} We have
\begin{equation}
\mu^{D}_{it} =
\frac{ \mu_{i0} }{ y_{it} } + \frac{ \xi \overline{ \mu }_i }{ y_{it} }  \int_{0}^{t} y_{iu} du
+ \frac{1}{ \left( \sigma^{D}  \right)^2  y_{it} } \int_{0}^{t} \frac{ \nu_{iu} y_{iu}  }{ D_u } dD_u + \frac{ \sigma^{ \mu }  \phi_i }{  y_{it} }  \int_{0}^{t} y_{iu} ds_u ,
\label{1}
\end{equation}
where
\begin{equation}
y_{it} = \exp \left( \xi t +  \frac{1}{ ( \sigma^{D} )^2 } \int_{0}^{t} \nu_{is} ds  \right),
\label{change2}
\end{equation}
and the variance process
\[
\nu_{it} := E^{Q^{i}} \left[ \left( \mu^{D}_t - E^{Q^i} \left[ \mu^{D}_{t} \big| \mathcal{G}_{t} \right]
 \right)^2  \big| \mathcal{ G}_{t} \right]
\]
is deterministic and given by
\begin{equation}
\nu_{it} = \alpha_{i2}  ( \sigma^{D} )^2  \frac{  e^{ (
\alpha_{i2} - \alpha_{i1} )t } - 1 }{e^{ ( \alpha_{i2} - \alpha_{i1} )t } - \alpha_{i2} /
\alpha_{i1} },
\label{variance_dynamics}
\end{equation}
where
\[ \alpha_{i2} = \sqrt{ \xi^2 + ( \sigma^{ \mu} / \sigma^{D} )^2
( 1 - \phi^{2}_i ) } - \xi, \]
and
\[ \alpha_{i1} = - \sqrt{ \xi^2 +
( \sigma^{ \mu} / \sigma^{D} )^2 (  1 - \phi^{2}_i ) } -
\xi.
\]
\label{Lipster_Shiryayev}
\end{Proposition}

\textbf{Proof.}
Observe that Theorem 12.7 in Liptser and Shiryaev \cite{LS2} implies that $ \left( \mu^{D}_{it} \right)_{ t \in [0, \infty) }$ satisfies the following SDE
\begin{equation}
d \mu^{D}_{it} = - \xi \left( \mu^{D}_{it}  - \overline{ \mu }_i \right) dt + \frac{ \nu_{ it } }{ \left( \sigma^{D}  \right)^2  } \left( \frac{ d D_t }{ D_t } - \mu^{D}_{it} dt  \right)  +  \sigma^{ \mu } \phi_i ds_t,
\label{filter_dynamics}
\end{equation}
where the variance process $ \nu_{it} $ is detected through the following Riccati ODE
\[
\nu'_{it} = - 2 \xi \nu_{it} + ( \sigma_{ \mu })^2
\left( 1 - \phi^{2}_i \right)  - \frac{ 1  }{ ( \sigma^{D} )^2  }
\nu^{2}_{it},
\]
with $\nu_{i0} = 0.$ One can solve the above equation and verify that $ \nu_{it} $ is given by
$(\ref{variance_dynamics}).$ Now, we shall solve the SDE
(\ref{filter_dynamics}). By definition, we have
$ y'_{it} = ( \xi + \frac{ \nu_{it}  }{ (
\sigma^{D} )^2  }   ) y_{it} , $ and $ y_{ i 0 } = 1 $. Notice that the preceding observation combined with Ito's formula implies that
\[
d \left( y_{it} \mu^{D}_{it} \right) = \xi \overline{ \mu }_i y_{it} dt + \frac{ \nu_{it}  }{ \left( \sigma^{D}  \right)^2  } y_{it} \frac{ dD_t }{ D_t } +  \sigma^{ \mu } \phi_i y_{it} ds_t,
\]
completing the proof. \qed
${}$

\begin{Remark} Dumas, Kurshev and Uppal \cite{DKU} consider the static version of (\ref{1}).
That is, the functions $\nu_{it}$ and $y_{it}$ are substituted by the corresponding asymptotic limits.
This can be justified by Lemma \ref{estimates} of the current paper.
\end{Remark}

We denote by $i=0$ a fictional agent who is rational in the sense that he
knows the correct average, initial mean growth-rate and the correlation parameter $\phi$. Let us denote by
$\mu^{D}_{0t} := E^{P} \left[ \mu^{D}_{t}
\big| \mathcal{G}_t \right] $ the estimated mean growth-rate of this agent.
As in Proposition \ref{Lipster_Shiryayev}, we have
\[
\mu^{D}_{0t} =
\frac{ \mu_{0} }{ y_{0t} } + \frac{ \xi \overline{ \mu } }{ y_{0 t} }  \int_{0}^{t} y_{0u} du
+ \frac{1}{ \left( \sigma^{D}  \right)^2  y_{ 0 t} } \int_{0}^{t} \frac{ \nu_{0u} y_{0u}  }{ D_u } dD_u
+ \frac{ \sigma^{ \mu }  \phi }{  y_{0t} }  \int_{0}^{t} y_{0u} ds_u  ,
\]
where $y_{0t}$ and $\nu_{0t}$ are defined similarly to (\ref{change2}) and (\ref{variance_dynamics}). It can be shown, as in Theorem 8.1 in Liptser and Shiryaev \cite{LS1}, that
 $W^{(0)}_t = W^{(1)}_t - \int_{0}^{t} \frac{ \mu^{D}_{0s} - \mu^{D}_{s} }{ \sigma^{D} } ds $ is a $P-$Brownian motion with respect to the filtration $\mathcal{G}.$
 \begin{Remark}
 The filtration $\mathcal{G}$ is generated by the public signal $s_t$ and the
 Brownian motion $W^{(0)}_t$. To see this, note that $$ \frac{d D_t}{D_t} =
 \mu^{D}_{0t} dt + \sigma^{D} dW^{(0)}_t , $$ and
 $$ d \mu^{D}_{0t} = - \xi \left( \mu^{D}_{0t}  - \overline{ \mu } \right) dt + \frac{ \nu_{ 0t } }{ \sigma^{D}  }   dW^{(0)}_t +  \sigma^{ \mu } \phi ds_t.$$
 \label{filtration_rem}
 \end{Remark}
  We set
\begin{equation}
\delta_{it } := \frac{ \mu^{D}_{it} - \mu^{D}_{0t}  }{ \sigma^{D} }
\label{mgr_error}
\end{equation}
to be the $i-$th agent's \textit{error} in the mean growth-rate estimation.
The dynamics of $(D_t)_{ t \in [0, \infty) }$ from the $i-$th agent's perspective admit the form
$$ \frac{d D_t }{ D_t  } =  \mu^{D}_{it} dt + \sigma^{D} d {W}^{(0)}_{it} ,$$
where
\[
d {W}^{(0)}_{it} = d {W}^{(0)}_{t} - \delta_{it} dt
\]
is a Brownian motion (by Girsanov's theorem) under the equivalent probability
measure\footnote{One can check that the process $( Z_{it})_{t \in [ 0 , \infty)}  $ is a
true martingale by verifying Novikov's condition  on a small interval and then applying a similar argument
to the one used in Example 3, page 233, in Liptser and Shiryaev \cite{LS1}.} $P_{i}$ and the filtration $\mathcal{G}$,
where
\begin{equation}
Z_{it} := E \left[ \frac{ dP^{i}  }{ dP  } \big| \mathcal{G}_t \right] = \exp{ \left(   \int_{0}^{t} \delta_{is} dW^{(0)}_s - \frac{1}{2} \int_{0}^{t} \delta^{2}_{is}  ds\right)  }.
\label{A}
\end{equation}
Let us stress that ${W}^{(0)}_{it}$ is also a $Q^{i}-$Brownian motion
with respect to the filtration $\mathcal{G}.$ In particular, this implies
that by restricting the measure $Q^{i}$ to the sigma-algebra generated by
$W^{(0)}_{it}$, we get the measure $P^{i}.$
Nevertheless, the measures $Q^{i}$ and $P$ (the physical probability measure)
are singular on the sigma-algebra (see (\ref{change1}) and (\ref{signal_belief}))
generated by the Brownian motions $W^{(1)}_t, W^{(2)}_t$ and $W^{(3)}_t$.

\section{The Equilibrium State Price Density}

In the current section we depict the structure of the equilibrium
state price density in both settings of homogeneous and heterogeneous economies.

\subsection{Homogeneous Economy}
Consider an economy where all agents are of the same type $i$, and denote by
$(M_{it} )_{ t \in [ 0 ,\infty) } $ the corresponding equilibrium state price density.
The homogeneity of the economy combined with the completeness of the market
allows to derive the corresponding state price density in a closed form.

\begin{Lemma} The equilibrium state price density in a market populated by one agent of type $i$ is given by
\begin{equation}
M_{it} = e^{ - \rho_i t} D^{ - \gamma_i }_t Z_{it} H^{ \gamma_i - 1 }_{it} =
\label{SPD_h}
\end{equation}
\[
\exp \bigg(   - \int_{0}^{t} \bigg(  \rho_i  + \gamma_i \left( \mu^{D}_{0s} - \frac{1}{2} ( \sigma^D )^2 \right)  +  \frac{1}{2} \delta^{2}_{is} \bigg) ds \bigg) \times
\]
\[
\exp \left( \left( \gamma_i - 1  \right) \beta_i x_t + \int_{0}^{t} \left( \delta_{is} - \gamma_i \sigma^{D} \right) dW^{(0)}_s \right).
\]
\end{Lemma}
\textbf{Proof.} The assertion follows by using the
market clearing condition and Lemma
\ref{consumption_explicit}. \qed
${}$
\newline
We derive next the risk free-rate and the market price of risk in a homogeneous economy.
\begin{Lemma}
The risk free rate and the market price of risk in an economy populated by one agent of type $i$, are given respectively by
\[
r_{it} :=   \rho_i  + \gamma_i  \mu^{D}_{ it} - \frac{1}{2} \left( \sigma^{D} \right)^2 \gamma_i \left( \gamma_i + 1 \right)
-  \beta_i \left( \gamma_i - 1 \right)  \left( x_t - \log \left( D_t \right) \right)
\]
and
\[
\theta_{it} :=  \gamma_{i} \sigma^{D} - \delta_{it} .
\]
\label{homg_rdv}
\end{Lemma}
\textbf{Proof.} Consider the process
\[
Y_{it} := - \int_{0}^{t} (  \rho_i  + \gamma_i ( \mu^{D}_{0s} - \frac{1}{2} ( \sigma^D )^2 ) +  \frac{1}{2} \delta^{2}_{is} ) ds + \left( \gamma_i - 1  \right) \beta_i x_t + \int_{0}^{t} ( \delta_{is} - \gamma_i \sigma^{D}  ) dW^{(0)}_s.
\]
The dynamics of $M_{it}$ are given by
\[
\frac{ d M_{it}}{ M_{it} } =  dY_{it} + \frac{1}{2} d \langle Y_{i} , Y_{i} \rangle_{t} .
\]
where
\[
dY_{it} =  - \left( \rho_i  + \gamma_i \left( \mu^{D}_{t} - \frac{1}{2} ( \sigma^D )^2 \right) + \frac{1}{2} \delta^{2}_{it} \right) dt +
\]
\[
\beta_i ( \gamma_i - 1 ) (  \log(D_t)-x_t  ) dt + \left( \delta_{it} - \gamma_{i} \sigma^{D}  \right) dW^{(0)}_t ,
\]
and
\[
 d \langle Y_{i} , Y_{i} \rangle_{t} =  \left( \delta_{it} - \gamma_{i} \sigma^{D}  \right)^{2}  dt.
\]
The rest of the proof follows from the fact that the risk free rate and the market price of risk coincide with minus the drift and minus the volatility of the SPD respectively. $ \qed$

\subsection{Heterogeneous Economy}
Consider an economy populated by $N$ different types of agents. By Lemma \ref{consumption_explicit}, the optimal consumption
stream of agent $i$ is given by

\begin{equation}
c_{it} = e^{ - \frac{ \rho_i }{
\gamma_i } t }  M^{ - \frac{1 }{ \gamma_i } }_{t} Z^{
\frac{1}{ \gamma_i }}_{it}  H^{ \frac{ \gamma_i - 1 }{
\gamma_i } }_{it}  c_{i0} = c_{  i0 } \left( \frac{M_{it}  }{
M_t  }  \right)^{1 / \gamma_i }  D_t ,
\label{B}
\end{equation}
where $ ( M_t  )_{t \in [ 0 , \infty ) } $ stands for the corresponding heterogeneous equilibrium state price density, and $ M_{it} $ is given by
(\ref{SPD_h}). Therefore, the market clearing condition (\ref{MarketC}) admits the form
\begin{equation}
\sum_{i=1}^{N} c_{i0}  \left(  \frac{ M_{it}  }{ M_t  }  \right)^{ 1 / \gamma_i  } = 1.
\label{MC2}
\end{equation}

\begin{Example}
Consider a homogeneous risk aversion economy, i.e.,
$\gamma_1 = ... = \gamma_N = \gamma.$ Then, the equilibrium state price density is given explicitly by
\[
M_t = \left(  \sum_{i=1}^{N} \frac{ c_{i0} e^{ - \rho_i t / \gamma} Z^{ 1 / \gamma}_{it} H^{ \frac{ \gamma -1 }{ \gamma } } _{it}   }{ D_t }  \right)^{ \gamma} .
\]
Furthermore, if the habits are homogeneous, that is, $ \beta_1 =
... = \beta_N = \beta$, we have
\[
M_t =  e^{ ( \gamma -1 ) \beta x_t  } \left(  \sum_{i=1}^{N} \frac{ c_{i0} e^{ \rho_i t / \gamma} Z^{ 1 / \gamma}_{it} }{ D_t }  \right)^{ \gamma}.
\]
If the beliefs among the agents are not varying, i.e., $Z_{1t} = ... = Z_{Nt}  = Z_t $, then, we have
\[
M_t =  Z_{t} \left(  \sum_{i=1}^{N} \frac{ c_{i0} e^{ \rho_i t / \gamma} H^{ \frac{ \gamma -1 }{ \gamma } } _{it}   }{ D_t }  \right)^{ \gamma} .
\]
\label{C}
\end{Example}
Finally, we provide formulas for the risk free rate and the market price of risk.
\begin{Proposition} We have
\[
\theta_t = \sum_{i=1}^{N} \omega_{it} \theta_{it},
\]
and
\begin{align*}
r_t = \sum_{i=1}^{N} \omega_{it} r_{it} + \frac{1}{2} \sum_{i=1}^{N} ( 1 - 1 / \gamma_i ) \omega_{it} \left( \theta_{it} - \theta_{t} \right)^2,
\end{align*}
where
\begin{equation}
\omega_{it} := \frac{ 1 / \gamma_i c_{it} }{ \sum_{j=1}^{N} 1 / \gamma_j c_{jt} }
\end{equation}
denotes the relative level of absolute risk tolerance of agent $i$.
\label{interest_mpr}
\end{Proposition}
\textbf{Proof.} The proof is identical to the one of Proposition 4.1
in Cvitani\'c, Jouini, Malamud and Napp \cite{CJMN}. $ \qed$

\section{The Main Result: The Long-Run Surviving Consumer}
The current section is devoted to the study of the long-run
behavior of the optimal consumption shares in a heterogeneous
economy. We establish the existence of a surviving consumer in the
market, i.e., an agent whose optimal consumption asymptotically
behaves as the aggregate consumption. This dominating individual is
determined through the \textit{survival index}. The survival index is a
quantity depending on individuals' characteristics and
specifies the surviving agent versus the agents to be extinct in the
economy.

\begin{Definition}
The \textit{survival index} of agent $i$ is given by
\begin{equation}
\kappa_i := \rho_i + \left( \overline{\mu} - \frac{1}{2} ( \sigma^{D} )^2 \right) \left( \gamma_i + ( 1 - \gamma_i ) \beta_i \right) +
\label{sur_index_}
\end{equation}
\[
\frac{1}{2}  \left(  \frac{\overline{\mu}_i - \overline{\mu}}{\sigma^{D}}  \right)^2
 +
 \frac{ \xi^2 + \left( \sigma^{\mu} / \sigma^{D} \right)^2 \left( 1 - \phi \phi_i \right)  }{ 2 \sqrt{ \xi^2 +
 \left( \sigma^{\mu} / \sigma^{D} \right)^2 \left( 1 - \phi^2_{i} \right)  }  }   .
\]
\label{sur_index_definition}
\end{Definition}

The following is assumed throughout the entire paper.
\newline
\newline
\textbf{Assumption.}
There exists an agent $I_K$ whose survival index is
the lowest one, namely $ \kappa_{I_K} < \kappa_i,$ for all $ i \neq I_K.$
\newline
\newline
We are now ready to state our main result.

\begin{Theorem} In equilibrium, the only surviving agent in the long-run is the one with the lowest
survival index, i.e.,
\[
\lim_{t \to \infty } \frac{ c_{it} }{ D_t  } = 0
\]
for all $ i \neq I_{K} $, and
\[
\lim_{t \to \infty } \frac{ c_{I_{K}t} }{ D_t  } = 1.
\]
\label{surviving_consumer}
\end{Theorem}

The survival index is a complicated function of the individuals' underlying parameters. In order to isolate the effects of various agents'
characteristics on the long-run survival, we will discuss special cases in which
agents differ with respect to only one or a few particular parameters.

\subsection{The Effect of Risk-Aversion and Habits}
Let the initial priors $ \left( \overline{ \mu }_i \right)_{ i=1,...,N} $ and the over-confidence parameters $ \left( \phi_i \right)_{i=1,...,N} $ be fixed and identical for all agents. As it will be seen in the proof of Theorem \ref{surviving_consumer}, the survival index is invariant under additive translation, and thus it is determined in the current setting by
\[
 \rho_i + \left( \mu - \frac{ ( \sigma^{D}  )^2 }{2} \right)  (  \gamma_i + ( 1 - \gamma_i  )  \beta_i ).
\]
If $\beta_1 = ... = \beta_N = 0,$ the survival index is the same as in Cvitani\'c, Jouini, Malamud and Napp \cite{CJMN}. In particular, in a growing economy (i.e. $\mu -
(\sigma^{D} )^2 / 2 > 0 $), the least risk-averse agent will
survive in the long-run, as in the models of Yan \cite{Y1}, and Cvitani\'c, Jouini, Malamud and Napp \cite{CJMN}.
The presence of habits may change the behavior. Here, if the habit is
sufficiently strong ($\beta_i > 1$), the effect completely reverses: It is the most risk-averse agent who survives in the long-run. Effectively, 'catching up with Joneses'
preferences change an agent's risk aversion from $\gamma_i$ to
\begin{equation}
\gamma_i + ( 1 - \gamma_i ) \beta_i .
\label{effective_ra}
\end{equation}
Therefore, for strong habits, agents with a high risk-aversion
effectively behave as agents with a low risk aversion.
When risk aversion is homogeneous, the effect of habits strength on survival depends on whether risk aversion is above or below 1.
If risk aversion is above 1, we get the surprising, and at first sight counter-intuitive result, that agents with stronger habits
survive in the long-run.
The reason for this is that the presence of habits forces the
agent to trade more aggressively and make bets on very good
realizations of the dividend in order to sustain the aggregate
habit level generated by the 'catching up with the Joneses'
preferences. This makes an agent with strong habits effectively
less risk averse. This is beneficial for survival in a growing
economy.

\subsection{The Effect of Diverse Beliefs}
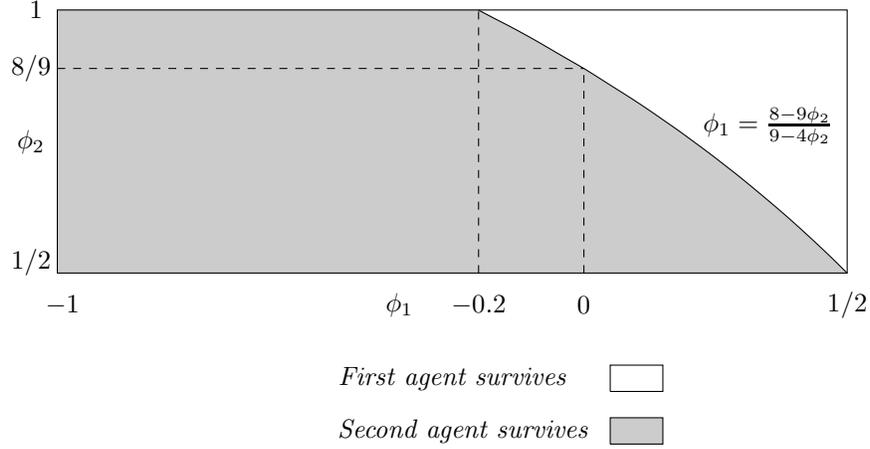
\begin{figure}[h]
\begin{center}
\begin{tikzpicture}[scale=7]

\filldraw[draw=black!20,fill=black!20,smooth] (0, 0.5) -- (0,8/9) -- (0.5, 0.5) -- (0,0.5);
\filldraw[draw=black!20,fill=black!20,smooth] (0, 0.5) -- (0, 8/9) -- (-0.2, 1) -- (-0.2,0.5) -- (0,0.5);
\filldraw[draw=black!20,fill=black!20,smooth] (-1, 0.5) -- (-1,1) -- (-0.2, 1) -- (-0.2,0.5);

\node (v00x) at (0,0.44) [circle,draw=blue!0,fill=blue!0,thick,inner sep=0pt,minimum size=2mm] {$0$};
\node (v10x) at (0.5,0.44) [circle,draw=blue!0,fill=blue!0,thick,inner sep=0pt,minimum size=2mm] {$1/2$};
\node (phi1) at (-0.35,0.44) [circle,draw=blue!0,fill=blue!0,thick,inner sep=0pt,minimum size=2mm] {$\phi_1$};

\node (v00y) at (-1.05,0.52) [circle,draw=blue!0,fill=blue!0,thick,inner sep=0pt,minimum size=2mm] {$1/2$};
\node (v01y) at (-1.04,1) [circle,draw=blue!0,fill=blue!0,thick,inner sep=0pt,minimum size=2mm]
{$1$};
\node (phi2) at (-1.05,0.75) [circle,draw=blue!0,fill=blue!0,thick,inner sep=0pt,minimum size=2mm] {$\phi_2$};
\node (-1) at (-0.99,0.44) [circle,draw=blue!0,fill=blue!0,thick,inner sep=0pt,minimum size=2mm]
{$-1$};

\node (-0.2) at (-0.2,0.44) [circle,draw=black!0,fill=black!0,thick,inner sep=0pt,minimum size=2mm] {$-0.2$};

\node (8/9) at (-1.05,8/9) [circle,draw=black!0,fill=black!0,thick,inner sep=0pt,minimum size=2mm] {$8/9$};

\node (phi2) at (-0.25, 0.30) [draw=blue!0,fill=blue!0,thick,inner sep=1pt,minimum size=1mm]
{\it{First agent survives}};

\draw (0.05, 0.275) -- (0.05,0.325) -- (0.15, 0.325) -- (0.15, 0.275) -- (0.05,0.275);
\filldraw[draw=black!100,fill=black!20]
(0.05, 0.175) -- (0.05,0.225) -- (0.15, 0.225) -- (0.15, 0.175) -- (0.05,0.175);

\node (phi2) at (-0.23 , 0.20) [draw=blue!0,fill=blue!0,thick,inner sep=1pt,minimum size=1mm] {\it{Second agent survives}};

\node (function) at (0.35, 0.78) [circle,draw=blue!0,fill=blue!0,thick,inner sep=1pt,minimum size=1mm] {$\phi_1 = \frac{8-9 \phi_2 }{ 9 - 4 \phi_2 }$};

\draw (-1,0.5) -- (0.5,0.5) -- (0.5,1) -- (-1,1) -- (-1,0.5);


\filldraw[draw=black!100,fill=black!20,scale=1,domain=-0.2:0.5,smooth,variable=\t]
plot ({\t},{ (9*\t-8)*(4*\t-9)^(-1) });

\draw[dashed] (0,0.5) -- (0,8/9);
\draw[dashed] (-1,8/9) -- (0,8/9);
\draw[dashed] (-0.2,0.5) -- (-0.2,1);

\end{tikzpicture}
\caption{The long-run surviving consumer}\label{fig:surv}
\end{center}
\end{figure}

Consider an economy where agents may differ only with respect to their average mean growth-rate estimations $( \overline{\mu}_i)_{i=1,...,N}$ and their correlation parameters $( \phi_i)_{i=1,...,N}$.
In this case, the survival index admits the form
\begin{equation}
\kappa_i =
\frac{1}{2}  \left(  \frac{\overline{\mu}_i - \overline{\mu}}{\sigma^{D}}  \right)^2
 +
 \frac{ \xi^2 + \left( \sigma^{\mu} / \sigma^{D} \right)^2 \left( 1 - \phi \phi_i \right)  }{ 2 \sqrt{ \xi^2 +
 \left( \sigma^{\mu} / \sigma^{D} \right)^2 \left( 1 - \phi^2_{i} \right)  }  }   .
\label{sur_index_2}
\end{equation}
Note that in this case the survival index is a decreasing function of the
correlation parameter $\phi_i$ in the interval $ [ -1, \phi],$ and an increasing function
in the interval $(\phi,1]$. Therefore, in an economy where the only distinction between agents
is their correlation parameters, the surviving agent is derived as follows. If either all agents are over-confident
($\phi < \phi_i,$ for all $i=1,...,N$) or under-confident ($\phi > \phi_i,$ for all $i=1,...,N$), then the survival index is given by
\[
\left|  \phi_i - \phi \right| ,
\]
and thus, the individual with the most accurate guess of the correct correlation will dominate the market. If some agents are
over-confident and some are under-confident in the signal, the situation
becomes more complex. For simplicity, let us analyze the case of an economy which consists of two agents: the first agent
underestimates the correlation and believes that it is
$ \phi_1 \in [ -1 , \phi ] $, whereas the second agent overestimates the correlation by $ \phi_2
\in [ \phi , 1] $. Let us set $a:= \left( \frac{ \xi \sigma^{D} }{ \sigma^{\mu} } \right)^2$.
If $\phi_{1} \in \left[ - 1 , 2 \frac{ a \phi \left( 1+a \right) }{ a \phi^2 + \left( a+1- \phi \right)^2 } -1 \right]$,
the second agent will survive. Now, assume that $ \phi_1 \in \left[
2 \frac{ a \phi \left( 1+a \right) }{ a \phi^2 + \left( a+1- \phi \right)^2 } - 1 , \phi \right]$. Then, if $\phi_{2} \in \left[ \phi ,
\frac{ 2 \left( a + 1 \right) \phi - \left(  a + 1 + \phi^2 \right) \phi_1 }{
 a + 1 + \phi^2 - 2 \phi \phi_1 }  \right],$ the second agent will survive; otherwise, namely, if
$\phi_{2} \in \left[ \frac{ 2 \left( a + 1 \right) \phi - \left(  a + 1 + \phi^2 \right) \phi_1 }{  a + 1 + \phi^2 - 2 \phi \phi_1 } , 1   \right],$
the first agent will survive.
To demonstrate the above scheme numerically, let us consider the case where $a=1$ and $\phi=1 / 2 $ (see Figure \ref{fig:surv}).
If $\phi_1 \in \left[ -1 , -0.2 \right] $, then the second agent will survive. If $ \phi_1 \in \left[ -0.2 , 0.5  \right] ,$ then: if
$\phi_2 \in \left[ 0.5 , \frac{ 8 - 9 \phi_1 }{ 9 - 4 \phi_1 }  \right] $, the second agent is the one to survive. Otherwise,
if $\phi_2 \in \left[ \frac{ 8 - 9 \phi_1 }{ 9 - 4 \phi_1 }  , 1 \right] $ then the first agent will survive. The preceding fact yields
an economically surprising observation: too overconfident agents will not survive when they compete with agents that believe in a weak
negative correlation. Assume, for instance, that the second agent believes that the correlation is some $\phi_2 \in [ 8/9 , 1 ]$.
Then, if $\phi_1 \in [ \frac{ 8 - 9 \phi_2 }{ 9 - 4 \phi_2 }  , 0] $, the first agent will survive, despite of the negative correlation.
This is very surprising, since irrational agents who believe in a non-positive correlation happen to survive, whereas individuals with an
overestimation of the signal will be extinct.

If the only source of heterogeneity in the economy is the belief regarding the average mean growth-rate, then the survival index depends only on the error between the subjective mean growth-rate and the correct one, namely,
 \[
\kappa_i = \left| \overline{\mu} - \overline{ \mu_i }  \right|.
\]
Therefore, the consumer with the best forecast of the average mean growth-rate is the one to dominate the market.

\subsection{The Relative Level of Absolute Risk Tolerance}
As in Cvitani\'c, Jouini, Malamud and Napp \cite{CJMN}, we define the \textit{relative level of absolute
risk tolerance} of agent $i$ by
\[
w_{ i t } := \frac{  1 / \gamma_i   c_{i t} }{   \sum_{ j = 1  }^{ N }  1 / \gamma_j    c_{j t}  }.
\]
The following is an immediate consequence of Theorem \ref{surviving_consumer}.
\begin{Corollary}
We have
\[
 \lim_{ t \to  \infty} w_{ i t } = 0
\]
for all $ i \neq I_k,$ and
\[
\lim_{ t \to  \infty} w_{ I_K t } = 1 .
\]
\label{tolerance}
\end{Corollary} \textbf{Proof.} Note that (\ref{B})
implies that
\[
w_{ i t } = \frac{c_{it}  }{D_t  }  \frac{ 1 / \gamma_i }{
\sum_{j=1 }^{N } 1 / \gamma_j  c_{ 0 j  } ( M_{jt} / M_t )^{
1 / \gamma_j }  }.
\]
The identity (\ref{MC2}) yields
\[
\frac{1}{ \sum_{j=1 }^{N } \frac{1}{ \gamma_j} c_{ 0 j  } ( M_{jt} / M_t )^{ 1 / \gamma_j }  } \leq \max_{k=1,...,N } \gamma_k.
\]
The preceding observations combined with Theorem \ref{surviving_consumer} and the equality $\sum_{i=1}^{N} \omega_{it} = 1 $ complete the proof of Corollary \ref{tolerance}. $\qed$

\section{Auxiliary Results}
In the present section we provide some results that will be crucial for proving Theorem \ref{surviving_consumer}.
First, we introduce the following estimates, indicating that
$y_{it},$ $1 / y_{it}$, their derivatives, and $\nu_{it}$ are close
to certain functions, of a simpler form. The errors in these estimates are shown to be decaying  exponentially fast to 0, as $t \to \infty$.
\begin{Lemma} We have
\newline
\begin{equation}
\left| \nu_{it} -  \alpha_{i2} ( \sigma^{D})^2    \right| \leq C e^{
- 2 ( \alpha_{i2} + \xi  ) t } , \label{estimate1}
\end{equation}

\begin{equation}
\left| y_{it} - \exp \left( - \frac{ \alpha_{i2}  }{ \alpha_{i1}  } e^{ -
\frac{ \alpha_{i2}  }{ \alpha_{i1}  }  }    \right) e^{ ( \alpha_{i2}  +  \xi) t
}   \right| \leq C  e^{  - \left( \alpha_{i2} + \xi  \right) t   } ,
\label{estimate2}
\end{equation}

\begin{equation}
\left| y'_{it} - ( \alpha_{i2}  +  \xi)  \exp \left( - \frac{ \alpha_{i2}
}{ \alpha_{i1}  } e^{ - \frac{ \alpha_{i2}  }{ \alpha_{i1}  }  }    \right)  e^{
( \alpha_{i2}  +  \xi) t  }  \right| \leq C e^{ - ( \alpha_{i2}  +  \xi) t
}  \label{estimate3}
\end{equation}
and

\begin{equation}
\left| \frac{1}{y_{it}} -  \exp \left(  \frac{ \alpha_{i2}  }{ \alpha_{i1}  } e^{ - \frac{ \alpha_{i2}  }{ \alpha_{i1}  }  }    \right)  e^{ - ( \alpha_{i2}  +  \xi )  t  }  \right| \leq C e^{ - 3 ( \alpha_{i2} + \xi )t  } ,
\label{estimate4}
\end{equation}

\begin{equation}
\left| \left( \frac{1}{y_{it}} \right)' + ( \alpha_{i2} + \xi )   \exp \left(  \frac{ \alpha_{i2}  }{ \alpha_{i1}  } e^{ - \frac{ \alpha_{i2}  }{ \alpha_{i1}  }  }    \right)  e^{ - ( \alpha_{i2}  +  \xi )  t  }  \right|
 \leq C  e^{ - 3 ( \alpha_{i2} + \xi ) t  } ,
\label{estimate5}
\end{equation}
for all $t>0$ and some constant $C>0$.
\label{estimates}
\end{Lemma}
${}$
\newline
\textbf{Proof.} Inequality (\ref{estimate1}) is due to the fact that
$
\left| \nu_{it} -  \alpha_{i2} ( \sigma^{D})^2    \right| =
\left| \frac{ ( \alpha_{i1} - \alpha_{i2} ) \alpha_{i2} ( \sigma^{D} )^2  }{ \alpha_{i1} e^{ 2 ( \alpha_{i2} + \xi ) t}  - \alpha_{i2}  } \right| .
$
Next, by definition (see Proposition \ref{Lipster_Shiryayev}), it follows that $y_{it}$ admits the form
\[
y_{it} = \exp \left( \left( \alpha_{i2}  + \xi \right) t - \frac{ \alpha_{i2} }{ \alpha_{i1} } e^{ - \frac{ \alpha_{i2}  }{ \alpha_{i1}  } } \left( 1  - e^{ -  2 ( \alpha_{i2} +  \xi  )t }   \right)     \right).
\]
One checks that the inequality $ e^{x} - 1 \leq (e-1) x, $ for all
$0 \leq x \leq 1$ concludes the validity of (\ref{estimate2}).
Recall that $y_{it}$ satisfies the ODE $ y'_{it} = \left(
\xi + \frac{ \nu_{it} }{ ( \sigma^{D} )^2 }  \right) y_{it} $,
and thus we can estimate
\[
 \left| y'_{it} - ( \alpha_{i2}  +  \xi)  \exp \left( - \frac{ \alpha_{i2}  }{ \alpha_{i1}  } e^{ - \frac{ \alpha_{i2}  }{ \alpha_{i1}  }  }    \right)  e^{ ( \alpha_{i2}  +  \xi) t  }  \right| \leq
\]
\[
\exp \left( - \frac{ \alpha_{i2}  }{ \alpha_{i1}  } e^{ - \frac{ \alpha_{i2}  }{ \alpha_{i1}  }  }    \right)  e^{ ( \alpha_{i2}  +  \xi) t  }
\left|  \frac{ \nu_{it} }{ ( \sigma^{D} )^2 }  -  \alpha_{i2}  \right| +
\]
\[
\left( \xi + \frac{ \nu_{it}  }{ ( \sigma^{D}   )^2 } \right) \left| y_{it} - \exp \left( - \frac{ \alpha_{i2}  }{ \alpha_{i1}  } e^{ - \frac{ \alpha_{i2}  }{ \alpha_{i1}  }  }    \right)  e^{ ( \alpha_{i2}  +  \xi) t  } \right|,
\]
which implies (\ref{estimate3}) by applying inequalities (\ref{estimate1}) with (\ref{estimate2}). Inequalities (\ref{estimate4}) and (\ref{estimate5}) are proved in a similar manner. $\qed$
\newline
\newline
For each $d \geq 1$, we denote by $ \left( C_{0} \left( [0,1] ; \mathbb{R}^d \right) , || \cdot ||_{\infty}  \right)$ the space of all $\mathbb{R}^d$-valued continuous functions on the interval $ [0,1] $ vanishing at $0$ endowed with the sup topology.
\begin{Definition}
We denote by $ K^{(d)} $ the space of all functions $ f = (f_1,...,f_d) \in C_{0} \left( [0 , 1] ; \mathbb{R}^d \right)$, such that each component $f_i$ is absolutely continuous, and
\[
\sum_{i=1}^{d} \int_{0}^{T} (f_i'(x))^2 dx \leq 1 .
\]
\label{cameron-martin}
\end{Definition}

We note that $K^{(d)}$ is a compact subset of $C_{0} \left( [0,1] ; \mathbb{R}^d \right)$ (see Proposition 2.7, page 343, in Revuz and Yor \cite{RY}). The next result deals with the asymptotics of certain multiple stochastic integrals.

\begin{Lemma}
Let $(W_t)_{t \in [0 , \infty)} $ and $(B_t)_{t \in [0 , \infty)}$  be two arbitrary standard Brownian motions
and denote $ Z_t = \int_{0}^{t} e^{-s}  W_{ \frac{1}{2}  \left( e^{2s} - 1  \right) } d B_s.$
Then, we have
\newline
(i)
\[
\langle Z \rangle_{ \infty } := \lim_{ t \to \infty } \langle Z \rangle_t = \infty.
\]
(ii) \[
\lim_{t \to \infty} \frac{ \int_{0}^{t}  e^{-as} \int_{0}^{s} e^{au} dW_u   dB_s }{ t } = 0 ,
\]
for any $a>0.$
\newline
(iii)
\[
\lim_{t \to \infty} \frac{ \int_{0}^{t} e^{ - ( a + b ) s } \int_{0}^{s}  e^{  a  u } \int_{0}^{u} e^{ b x } dW_{x}   du dB_s  }{t } = 0,
\]
for all $a,b>0$.
\label{strassen}
\end{Lemma}
${}$
\newline
\textbf{Proof.}
(i) First, note that a change of variable implies that
$
\langle Z \rangle_t  =  \int_{0}^{t} e^{- 2s}  \left( W_{ \frac{1}{2}  \left( e^{2s} - 1  \right) } \right)^2 ds =
 \int_{0}^{ \frac{1}{2} ( e^{2t} - 1 )  } \frac{ W^2_{ u }  }{ (1 + u)^2 } du .
$
Consider the functional $F : C_{0} \left( [0,1]; \mathbb{R} \right)  \to \mathbb{R}_{+}$, which is given by
\[
F ( f) : = \int_{0}^{1} \frac{ f^2(x) }{ ( 1 + x)^2  } dx.
\]
Note that $F$ is a continuous functional. Indeed, for a fixed $f \in C_{0} \left( [0,1]; \mathbb{R} \right) $ and all $ \varepsilon > 0$,
let $ \delta = \varepsilon (2 || f ||_{ \infty } + \varepsilon  ) $ and observe that $|| f - g ||_{ \infty } < \delta  $
for some $g \in C_{0} \left( [0,1]; \mathbb{R} \right) $, implies
$ | F(f) - F(g) | < \varepsilon.$ It follows by Strassen's functional law of iterated logarithm
(see Theorem 2.12, page 346, in Revuz and Yor \cite{RY}) that P-a.s
\[
\limsup_{ N \to \infty } F \left( \frac{1}{ \sqrt{ 2 N \log \log (N)} } W_{Nt} \right) = \sup_{  h \in K^{(1)} } F( h ) .
\]
Notice that $ \sup_{ h \in K^{(1)} } F( h )  \geq F( \tilde{h} ) > 0, $ where $\tilde{h}(x) = x .$ Therefore, we have
\[
\limsup_{ N \to \infty } F \left( \frac{1}{ \sqrt{ 2 N \log \log (N)} } W_{Nt} \right)
\]
\[
= \limsup_{ N \to \infty} \frac{ \int_{0}^{1} \frac{ W_{Nt} }{ ( 1 +t)^2 } dt}{ 2 N \log \log N} =  \limsup_{ N \to \infty} \frac{ \int_{0}^{N} \frac{ W^2_u }{ ( N + u  )^2 }   du }{ 2 \log \log (N)}
> 0.
\]
Furthermore,
\[
 \limsup_{ N \to \infty} \frac{ \int_{0}^{N} \frac{ W^2_u }{ ( 1 + u  )^2 }   du }{ 2 \log \log (N)}
\geq
\limsup_{ N \to \infty} \frac{ \int_{0}^{N} \frac{ W^2_u }{ ( N + u  )^2 }   du }{ 2 \log \log (N)} > 0.
\]
In particular, it follows that $ \limsup_{ N \to \infty} \int_{0}^{N} \frac{ W^2_u }{ ( 1 + u  )^2 }   du = \infty ,$
but, since the function $ \int_{0}^{N} \frac{ W^2_u }{ ( 1 + u  )^2 }   du $ is monotone increasing in $N$, it follows that
$ \lim_{ N \to \infty} \int_{0}^{N} \frac{ W^2_u }{ ( 1 + u  )^2 }   du = \infty $. This accomplishes the proof of part (i).
\newline
(ii) Denote $ Y_t =  \int_{0}^{t}  e^{-s} \int_{0}^{s} e^{u} dW_u   dB_s $
and $ X_s = \int_{0}^{s} e^{u} dW_u $. Note that
$ \left\langle X  \right\rangle_t = \frac{1}{2}  \left( e^{2t} - 1  \right)  $. Since
$X_t$ is a martingale vanishing at 0 and $ \langle X \rangle_{ \infty} = \infty, $ it follows
by the Dambis, Dubins-Schwartz theorem (shortly DDS, see Theorem 1.6, page 181, in Revuz and Yor \cite{RY}) that
$ X_t = \widetilde{W}_{ \frac{1}{2}  \left( e^{2t} - 1  \right) },$
for a certain Brownian motion $\widetilde{W}_t.$ Therefore, we can rewrite
$$ Y_t = \int_{0}^{t} e^{-s}  \widetilde{W}_{ \frac{1}{2}  \left( e^{2s} - 1  \right) } d B_s, $$
and thus by part (i), we have $ \lim_{t \to \infty } \langle Y \rangle_t  = \langle Y \rangle_{ \infty} = \infty.$
It follows from the DDS theorem
that $Y_t = \widetilde{B}_{ \langle Y \rangle_t } $, for some
Brownian motion $\widetilde{B}_t.$ Now, denote $ \phi(x) = \sqrt{2 x \log \log x}$ and rewrite $
\frac{ Y_t  }{ t } = \frac{ \widetilde{B}_{ \langle Y \rangle_t }  }{ \phi ( \langle Y \rangle_t ) }   \frac{ \phi \left( \langle Y \rangle_t \right)}{ t }.
$
By the law of iterated logarithm, we have
$ \limsup_{ t \to \infty} \frac{ | \widetilde{B}_{ \langle Y \rangle_t }  | }{ \phi ( \langle Y \rangle_t ) }  \leq 1 $,
and hence it is enough to concentrate on the asymptotics of the second term:
\[
\frac{ \phi \left( \langle Y \rangle_t \right)}{ t } = \sqrt{ 2  \frac{ \int_{0}^{t} e^{- 2s}  \left( \widetilde{W}_{ \frac{1}{2}  \left( e^{2t} - 1  \right) } \right)^2 ds  \log \log \left( \int_{0}^{t} e^{- 2s}  \left( \widetilde{W}_{ \frac{1}{2}  \left( e^{2t} - 1  \right) } \right)^2 ds \right)    }{ t^2 } } .
\]
Note that $\phi(  \frac{1}{2} ( e^{2s} - 1 )   ) \leq e^{s} \sqrt{ \log 2s } $
and thus, the law of iterated logarithm implies that
\[
\limsup_{ t \to \infty} \frac{ \phi \left( \langle Y \rangle_t \right)}{ t } \leq \limsup_{ t \to \infty} \sqrt{ \frac{ \log(2t) \log \log (t \log 2t) }{ t } }  = 0.
\]
 This accomplishes the proof of part (ii).
\newline
(iii) By Fubini's theorem, we have
\[
\lim_{t \to \infty} \frac{ \int_{0}^{t} e^{ - ( a + b ) s } \int_{0}^{s}  e^{  a  u } \int_{0}^{u} e^{ b x } dW_{x}   du dB_s  }{t } =
\]
\[
\frac{1}{a}
\lim_{t \to \infty} \frac{ \int_{0}^{t}  e^{-a s } \int_{0}^{s} e^{bx} d W_x d B_s    }{ t }
- \frac{1}{a}
\lim_{t \to \infty} \frac{ \int_{0}^{t}  e^{- ( a + b )  s } \int_{0}^{s} e^{ (a+b) x} d W_x d B_s    }{ t } = 0 ,
\]
where the last equality follows by part (ii).
This completes the proof of Lemma \ref{strassen}. $\qed$
\newline
\newline
We proceed with the following statement.
\begin{Lemma} Let $(W_t)_{t \in [0 , \infty)} $ be a standard Brownian motion. Then, we have
\newline
(i)
\[
\lim_{ t \to \infty } \frac{ \int_{0}^{t} e^{-as} \int_{0}^{s} e^{ax} dW_x ds  }{t} = 0
\]
for all $a>0$.
\newline
(ii)
\[
\lim_{t \to \infty} \frac{ \int_{0}^{t} e^{ - ( a + b ) s } \int_{0}^{s} e^{au} \int_{0}^{u} e^{bx} dW_x du ds  }{ t } = 0
\]
for all $a,b>0$.
\label{no_stationary}
\end{Lemma}
${}$
\newline
\textbf{Proof.} (i) By using integration by parts and Fubini's theorem, we get
\[
\lim_{ t \to \infty} \frac{ \int_{0}^{t} e^{ - a  s   }  \int_{0}^{s} e^{ a  u  } d W_{u}   ds  }{t} =
\lim_{ t \to \infty} \frac{ \int_{0}^{t} \left( W_s -   a  e^{ - a s}  \int_{0}^{s} e^{ a u} W_u  du \right)  ds }{t} =
\]
\[
\lim_{t \to \infty} \frac{ \int_{0}^{t} W_s ds - a  \int_{0}^{t} \left( e^{ a u} W_u  \int_{u}^{t} e^{ - a s}  ds \right)   du    }{t} =
\lim_{t \to \infty} \frac{  \int_{0}^{t} e^{ a u} W_u  du    }{t e^{a t} } = 0,
\]
where the last equality follows by the law of large numbers.
\newline
(ii) As in (i), one checks that the limit is equal to
\[
\frac{1}{a} \lim_{  t \to \infty} \left( \int_{0}^{t} e^{-bs} \int_{0}^{s} e^{bx} dW_x ds -
\int_{0}^{t}  e^{- ( a + b) s} \int_{0}^{s} e^{ ( a  + b) x} dW_x ds     \right),
\]
which vanishes according to (i). $\qed$
\newline
\newline
In the next limit theorems, the main tool is ergodicity of certain stochastic processes. Similar ideas as below
(even though we have provided a direct argument) could be
applicable to deduce the previous lemma.
\begin{Lemma}
Let $ \left( W_t \right)_{t \in [0 , \infty ) } $ and $ \left( B_t \right)_{t \in [0 , \infty ) } $ be two independent Brownian motions.
Then, the following holds
\newline
(i)
\[
\lim_{t \to \infty} \frac{  \int_{0}^{t} e^{ - a s }  \int_{0}^{s} e^{ax} dW_x e^{ - b s }  \int_{0}^{s} e^{bx} dB_x  ds }{ t } = 0
\]
for all $a,b>0$.
\newline
(ii)
\[
\lim_{t \to \infty } \frac{ \int_{0}^{t} \left(  e^{-as} \int_{0}^{s} e^{ax} dW_x  \right)^2 ds }{ t }  = \frac{1}{2a}
\]
for all $a>0$.
\newline
(iii)
\[
\lim_{t \to \infty } \frac{ \int_{0}^{t}   e^{-(a + b ) s} \int_{0}^{s} e^{ax} dW_x  \int_{0}^{s} e^{bx} dW_x   ds }{ t }  = \frac{1}{a  + b }
\]
for all $a, b >0$.
\label{stationary}
\end{Lemma}
${}$
\newline
\textbf{Proof.} (i) First observe that $\int_{0}^{s} e^{ax} dW_x $  is a martingale with
$ \langle \int_{0}^{ \cdot } e^{ax} dW_x \rangle_t = \frac{ e^{ 2 a t }  - 1 }{ 2a}  $, and thus by the DDS theorem, we have
$ \int_{0}^{t} e^{ax} dW_x  = \widetilde{W}_{ \frac{ e^{ 2at } - 1 }{ 2 a } } $ for some Brownian motion
$ ( \widetilde{W}_t )_{t \in [0 , \infty ) } $. A similar argument implies that $ \int_{0}^{t} e^{bx} dB_x  = \widetilde{B}_{ \frac{ e^{ 2bt } - 1 }{ 2 b } } $
for a Brownian motion $ ( \widetilde{B}_t )_{t \in [0 , \infty ) } $. The construction in the DDS theorem implies that
$( \widetilde{B}_t )_{t \in [0 , \infty ) }$ and $( \widetilde{B}_t )_{t \in [0 , \infty ) }$ are independent.
Recall that $ e^{ - a t } \widetilde{W}_{  e^{ 2at}  }$ and $ e^{ - b t } \widetilde{B}_{ e^{ 2bt}  } $ are two independent stationary
Ornstein-Uhlenbeck processes, thus the process $ e^{ - (a + b) t } \widetilde{W}_{  e^{ 2at}  } \widetilde{B}_{  e^{ 2bt}  } $ is stationary.
Therefore, an ergodic theorem for stationary processes implies that
\begin{equation}
\lim_{t \to \infty} \frac{ \int_{0}^{t} e^{ - (a + b) s } \widetilde{W}_{  e^{ 2as}  } \widetilde{B}_{  e^{ 2bs}  } ds  }{t} =
e^{ - (a + b)  } E \left[  \widetilde{W}_{  e^{ 2a}  } \widetilde{B}_{  e^{ 2b}  }  \right] = 0.
\label{ergodic1}
\end{equation}
Next, the process $ ( W'_t )_{ t \in [0, \infty) } $ given by $W'_t = \sqrt{2a} \widetilde{W}_{\frac{t}{2a}}$ for $t < 1,$ and $W'_t = \sqrt{2a} \widetilde{W}_{\frac{t-1}{2a}} + \sqrt{2a} \widetilde{W}_{ \frac{1}{2a}} $ for $t > 1$ is a Brownian motion. Thus, we have $ \widetilde{W}_{ \frac{ e^{ 2as } -1 }{ 2a }  }  =
\frac{1}{\sqrt{2a}} W'_{ e^{ 2as } } - \widetilde{W}_{ \frac{1}{2a}},$ for all $s>1$. We define the process $(B'_t)_{ t \in [ 0 , \infty)}$ in a similar manner. We emphasize that
$ ( W'_t )_{ t \in [0, \infty) } $ and $ (\widetilde{W}_t )_{ t \in [0, \infty) } $ are independent of $ ( B'_t )_{ t \in [0, \infty) } $ and $ (\widetilde{B}_t )_{ t \in [0, \infty) } $. Thus we can rewrite (\ref{ergodic1}) as
\[
\lim_{t \to \infty} \frac{  \int_{0}^{t} e^{ - (a+b) s }  \left( \frac{1}{ \sqrt{2a}} W'_{ e^{ 2as } } - \widetilde{W}_{ \frac{1}{2a} }  \right) \left( \frac{1}{ \sqrt{2b}} B'_{ e^{ 2bs } } - \widetilde{B}_{ \frac{1}{2a} }  \right)   ds }{ t }.
\]
Next, the law of iterated logarithm implies that for every $ \varepsilon > 0 $ there exists an $ \mathcal{F}_{ \infty}$-measurable random variable
$N( \varepsilon ) : \Omega \to \mathbb{R}_{+} $ such that for all $ s > N ( \varepsilon)$, $ \left| \frac{ W_{ e^{2as} } }{ e^{ as \sqrt{ \log (2as) }  }}  \right| < 1 + \varepsilon ,$ and hence
\[
\lim_{t \to \infty} \frac{ \int_{0}^{t} \left| e^{ - as - bs } W'_{ e^{2as} }  \right| ds }{  t } \leq (1 + \varepsilon) \lim_{t \to \infty} \frac{ \int_{0}^{t} \frac{ \log ( as) }{ e^{bs}}  ds }{ t }  = 0.
\]
This fact combined with (\ref{ergodic1}) accomplishes the proof of part (i).
\newline
(ii) As in (i),  $  \int_{0}^{s} e^{ax} dW_x = \widetilde{W}_{ \frac{ e^{2as} - 1 }{ 2a } } $ and
$\widetilde{W}_{ \frac{ e^{ 2as } -1 }{ 2a }  }  = \frac{1}{\sqrt{2a}} W'_{ e^{ 2as } } - \widetilde{W}_{ \frac{1}{2a}}$.
Next, ergodicity yields
\[
\lim_{t \to \infty} \frac{ \int_{ 0 }^{t} \left(  e^{-as} \widetilde{W}_{ e^{ 2as} }  \right)^2 ds  }{ t } = \frac{1}{ e^{2a} } E \left[ \widetilde{W}^2_{ e^{2a}}  \right] = 1.
\]
Finally, the above limit combined with similar arguments to those appearing in (i) concludes the proof.
\newline
(iii) The idea of the proof is to rewrite the required limit in terms of limits of the same form as those in (ii). First, observe that $ e^{ - a t } \int_{0}^{s} e^{au} dW_u = W_s - a e^{-at} \int_{0}^{s} e^{au} W_u du. $
Thus we can rewrite,
\begin{equation}
 \int_{0}^{t} \left( e^{ - a s } \int_{0}^{s} e^{au} dW_u \right)^2 ds
\label{square}
\end{equation}
\[
= \int_{0}^{t} W^2_s ds  - 2a  \int_{0}^{t} e^{-as} W_s \int_{0}^{s} e^{au} W_u du ds + a^2 \int_{0}^{t} e^{- 2as} \left( \int_{0}^{s} e^{au} W_u du \right)^2 du.
\]
Observe that Fubini's theorem implies that
\begin{equation}
\int_{0}^{t} e^{- 2as} \left( \int_{0}^{s} e^{au} W_u du \right)^2 du = \int_{0}^{t} \int_{0}^{t} e^{ ax + ay} W_x W_y  \int_{ \max \{ x , y  \} }^{t}  e^{-2as} ds dx dy
\label{square2}
\end{equation}
\[
= \frac{1}{a} \int_{0}^{t} e^{-as} W_s \int_{0}^{s} e^{au} W_u du ds - \frac{1}{2a e^{2at} } \left( \int_{0}^{t} W_x e^{ax} dx \right)^2.
\]
This fact alongside (\ref{square}) and (\ref{square2}) implies that
\[
\lim_{ t \to \infty} \frac{ \int_{0}^{t} \left( e^{ - a t } \int_{0}^{s} e^{au} dW_u \right)^2 ds}{t}
\]
\[
= \lim_{ t \to \infty} \frac{ \int_{0}^{t} W^2_s ds - a  \int_{0}^{t} e^{-as} W_s \int_{0}^{s} e^{au} W_u du ds - \frac{a}{2 e^{ 2 a t } } \left( \int_{0}^{t} e^{as} W_s ds \right)^2  }{t}.
\]
By using similar arguments and exploiting the preceding observations, one can check that
\begin{equation}
\lim_{t \to \infty} \frac{ \int_{0}^{t}   e^{-(a + b ) s} \int_{0}^{s} e^{ax} dW_x  \int_{0}^{s} e^{bx} dW_x   ds }{t} =
\end{equation}
\[
\frac{a}{a+b} \lim_{t \to \infty } \frac{ \int_{0}^{t} W^2_s ds - a  \int_{0}^{t} e^{-as} W_s \int_{0}^{s} e^{au} W_u du ds + \frac{a}{2 e^{ 2 a t } } \left( \int_{0}^{t} e^{as} W_s ds \right)^2 }{ t } +
\]
\[
 \frac{b}{a+b} \lim_{t \to \infty } \frac{ \int_{0}^{t} W^2_s ds - b  \int_{0}^{t} e^{-as} W_s \int_{0}^{s} e^{au} W_u du ds  + \frac{a}{2 e^{ 2 a t } } \left( \int_{0}^{t} e^{as} W_s ds \right)^2 }{ t }.
\]
The latter fact combined with part (ii) completes the proof. $\qed$
\newline
\newline
The next statement is heavily based on the previous lemma.
\begin{Lemma}
Let $ \left( W_t \right)_{t \in [0 , \infty ) } $ and $ \left( B_t \right)_{t \in [0 , \infty ) } $ be two independent Brownian motions.
Then, we have
\newline
(i)
\[
\lim_ {t \to \infty} \frac{ \int_{0}^{t}  \left( e^{ - ( a + b ) s } \int_{0}^{s} e^{ax} \int_{0}^{x} e^{ b u} dW_u dx   \right)^2 ds   }{t}
= \frac{ 1   }{ 2 b ( a + b) (  a + 2 b ) }
\]
for all $a,b>0$.
\newline
(ii)
\[
\lim_{t \to \infty} \frac{ \int_{0}^{t} e^{ -( 2a + b )  s } \int_{0}^{s} e^{au} dW_u \int_{0}^{s} e^{bu} \int_{0}^{u} e^{ a x} dW_x du ds   }{t} = \frac{1}{ 2a ( 2a + b )}
\]
for all $a,b>0$.
\newline
(iii)
\[
\lim_{t \to \infty} \frac{ \int_{0}^{t} e^{ - (a + b ) s} \int_{0}^{s} e^{ (a-\xi) u } \int_{0}^{u} e^{\xi u} dW_{x} du \int_{0}^{s} e^{ (b-\xi) u } \int_{0}^{u} e^{\xi u} dW_{x} du ds  }{t} =
\]
\[
\frac{1}{ (a-\xi) (b-\xi) }  \left(  \frac{1}{a+b} + \frac{1}{2 \xi } - \frac{1}{ a + \xi } - \frac{1}{ b + \xi } \right)
\]
for all $a,b,\xi >0$.
\newline
(iv)
\[
\lim_{ t \to \infty} \frac{  \int_{0}^{t} e^{ -2 ( a + b ) s } \int_{0}^{s} e^{ay} \int_{0}^{y} e^{ b u } dW_u dy  \int_{0}^{s} e^{ (a+b)x } dW_x   ds }{ t }
= \frac{1}{2 (a + b  ) ( a + 2b ) }
\]
for all $a,b,\xi >0$.
\newline
(iv)
\[
\lim_{ t \to \infty} \frac{  \int_{0}^{t} e^{ -2 ( a + b ) s } \int_{0}^{s} e^{ay} \int_{0}^{y} e^{ b u } dW_u dy  \int_{0}^{s} e^{ (a+b)x } dW_x   ds }{ t }
= \frac{1}{2 (a + b  ) ( a + 2b ) }
\]
for all $a,b>0$.
\newline
(v)
\[
\lim_{ t \to \infty} \frac{  \int_{0}^{t} e^{ -2 ( a + b ) s } \int_{0}^{s} e^{ay} \int_{0}^{y} e^{ b u } dW_u dy  \int_{0}^{s} e^{ (a+b)x } dB_x   ds }{ t } =0
\]
for all $a,b>0$.
\label{last}
\end{Lemma}

\textbf{Proof.}
(i) Notice that  $ \int_{0}^{s} e^{ax} \int_{0}^{x} e^{ b u} dW_u dx =
\frac{1}{a} \int_{0}^{s} e^{bu} ( e^{as} - e^{au} )  dW_u.$ Therefore, the required limit is equal to
\[
\frac{1}{ a^2 } \lim_{ t \to \infty} \frac{ \int_{0}^{t} \left( e^{-bs} \int_{0}^{s} e^{bu} dW_u  \right)^2 ds   }{ t }
   - \frac{2}{ a^2 } \lim_{ t \to \infty} \frac{ \int_{0}^{t}  e^{- ( a + 2 b) s} \int_{0}^{s} e^{ ( a + b) u} dW_u \int_{0}^{s} e^{bu} dW_u   ds   }{ t }
\]
\[
+ \frac{1}{ a^2 } \lim_{ t \to \infty} \frac{ \int_{0}^{t} \left( e^{- ( a + b) s} \int_{0}^{s} e^{ ( a + b) u} dW_u  \right)^2 ds   }{ t } .
\]
Parts (ii) and (iii) in Lemma \ref{stationary} complete the proof of (i).
\newline
(ii) As before, one checks that the limit is equal to
\[
\frac{1}{b} \lim_{t \to \infty} \frac{ \int_{0}^{t}  e^{-2as} \left(  \int_{0}^{s} e^{ax} dW_x \right)^2 ds  }{t}
\]
\[
- \frac{1}{b} \lim_{t \to \infty} \frac{ \int_{0}^{t} e^{ - ( 2a + b ) s } \int_{0}^{s} e^{au} dW_u \int_{0}^{s} e^{ ( a+b) x  } dW_x ds   }{t} ,
\]
and the rest is a consequence of parts (ii) and (iii) of Lemma \ref{stationary}.
\newline
(iii) The limit is equal to
\[
\frac{1}{ ( a - \xi ) ( b - \xi ) } \bigg( \lim_{t \to \infty } \frac{ \int_{0}^{t}
\left( e^{-au} \int_{0}^{u} e^{ax} dW_x \right)^2 du +
\int_{0}^{t}  e^{-(a+b)u} \int_{0}^{u} e^{ax} dW_x \int_{0}^{u} e^{bx} dW_x   du   }{t}
\]
\[
- \lim_{t \to \infty } \frac{
\int_{0}^{t}  e^{-(a+ \xi )u} \int_{0}^{u} e^{ \xi x} dW_x \int_{0}^{u} e^{ax} dW_x   du
+ \int_{0}^{t}  e^{-( b + \xi )u} \int_{0}^{u} e^{\xi x} dW_x \int_{0}^{u} e^{bx} dW_x   du
}{t} \bigg).
\]
The rest follows by applying items (ii) and (iii) of Lemma \ref{stationary}.
\newline
(iv) One checks that the required limit is equal to
\[
\frac{1}{a} \lim_{ t \to \infty} \frac{ \int_{0}^{t} e^{ - ( 2a + b ) s} \int_{0}^{s} e^{bu} dW_u \int_{0}^{s} e^{( a + b ) x} dW_x    ds }{t}
+
\]
\[
\frac{1}{a} \lim_{ t \to \infty} \frac{ \int_{0}^{t} e^{ -   ( a + b ) s} \left( \int_{0}^{s} e^{ (a + b) u} dW_u \right)^2   ds }{t} ,
\]
and the follows by parts (ii) and (iii) of Lemma \ref{stationary}.
\newline
(v) As in (i), one checks that the limit is equal to
\[
\frac{1}{a} \lim_{ t \to \infty} \frac{ \int_{0}^{t} e^{ - ( 2a + b ) s} \int_{0}^{s} e^{bu} dW_u \int_{0}^{s} e^{( a + b ) x} dB_x    ds }{t}
-
\]
\[
\frac{1}{a} \lim_{ t \to \infty} \frac{ \int_{0}^{t} e^{ -  2 ( a + b ) s} \int_{0}^{s} e^{ (a + b) u} dW_u \int_{0}^{s} e^{ ( a + b ) x} dB_x    ds }{t} ,
\]
which vanishes due to part (i) of Lemma \ref{stationary}. $\qed$

\section{Proof of The Main Result}
We provide here a proof for Theorem \ref{surviving_consumer}. Fix an arbitrary $i \neq I_K$. Recall that $ \sum_{j=1}^{N} c_{jt} = D_t,$ and thus it suffices to show that
$ \lim_{t \to \infty } \frac{ c_{it} }{ D_t } = 0$.
Note that (\ref{MC2}) implies that $ M_{t} \geq c^{\gamma_{I_K}
}_{I_K 0 } \cdot M_{I_K t } $. Therefore, identity (\ref{B}) yields
\[
\frac{ c_{it } }{ D_t } = c_{0i}  \left( \frac{ M_{it}  }{ M_{t} }  \right)^{1 / \gamma_i }  \leq \frac{ c_{i0 } }{ c^{ \gamma_{I_K} / \gamma_i }_{ I_K 0 } }
\left( \frac{M_{it}  }{ M_{I_K t}  }  \right)^{1 / \gamma_i  }.
\]
In virtue of identity (\ref{SPD_h}), we have
\[
\frac{M_{it}  }{ M_{I_K t}} = \exp \left( a_{i}(t) - a_{I_K}(t) \right) ,
\]
where
\[
a_{j}(t) := ( \gamma_j - 1 ) \beta_j x_t + \left( \frac{ ( \sigma^{D} )^2 }{ 2 } \gamma_j - \rho_j  \right) t +
\]
\[
\int_{0}^{t} \left( - \gamma_j \mu^{D}_{s} - \frac{ \delta^{2}_{js}  }{2 } \right)  ds + \int_{0}^{t} \delta_{js}  d W^{(0)}_s
- \gamma_{ j } \sigma^{D} W^{(1)}_t ,
\]
for all $j=1,...,N.$ Therefore, in order to complete the proof of the statement, it suffices to show that
\[
\lim_{t \to \infty} \frac{ a_{i}(t)  - a_{I_K}(t) }{ t }  = \kappa_{ I_K } - \kappa_{i} < 0.
\]
To this end, we proceed with the computation of the following limits.
\newline
\newline
\textit{Part I.} We claim that
\begin{equation}
\lim_{t \to \infty}   \frac{x_t}{t}   =  \overline{\mu} - \frac{1}{2} (\sigma^{D})^2.
\label{x}
\end{equation}
Recall that by (\ref{standard_living}) and (\ref{divident_explicitly}), we have
\[
\lim_{t \to \infty} \frac{  x_t  }{ t  } =  \lim_{ t \to \infty} \frac{ x_0 +   \lambda  \int_{0}^{t} e^{ \lambda s}  \left( \int_{0}^{s} \mu_{u}^{D} du  - \frac{1}{2} ( \sigma^{D} )^{2} s + \sigma^{D} W^{(1)}_s \right) ds }{ t e^{ \lambda t  } } .
\]
Note that the law of large numbers implies that $ \lim_{ t \to \infty} \frac{ \int_{0}^{t} e^{ \lambda s } W^{(1)}_s ds   }{ t e^{ \lambda t} }  = 0.$
Next, it is evident that $  \lim_{ t \to \infty  } \frac{ x_0 }{ t e^{ \lambda t  } }  = 0  $  and $  \lim_{ t \to \infty  } \frac{ \int_{0}^{t} s e^{ \lambda s } ds }{ t e^{ \lambda t  } }  = 1 / \lambda.  $ Let us show that
\begin{equation}
\lim_{t \to \infty} \frac{ \int_{0}^{t} \mu^{D}_{u} du  }{ t  } =\overline{\mu}.
\label{mu2}
\end{equation}
By (\ref{mgr_explicit}), we get
\[
\lim_{ t \to \infty } \frac{ \int_{0}^{t} \mu^{D}_{u} du  }{ t  } = \lim_{ t \to \infty} \frac{ \int_{0}^{t}  \left(  \overline{\mu} + \left( \mu_0 - \overline{\mu} \right) e^{- \xi s}  + \sigma^{ \mu} \int_{0}^{s} e^{ \xi ( u - s  ) } d W^{(2)}_{u}  \right)   ds  }{t}.
\]
Clearly, we have $\lim_{ t \to \infty} \frac{ \int_{0}^{t}  \left(  \overline{\mu} + \left( \mu_0 - \overline{\mu} \right) e^{- \xi s} \right)   ds  }{t} = \overline{\mu} $. Furthermore, part (i) of Lemma \ref{no_stationary} yields $ \lim_{ t \to \infty} \frac{ \int_{0}^{t}   \int_{0}^{s} e^{ \xi ( u - s  ) } d W^{(2)}_{u}   ds  }{t} = 0.$ This asserts the validity of (\ref{mu2}). Next, by L'h\^opital's rule, we get
\[
\lim_{ t \to \infty }  \frac{ \int_{0}^{t} e^{ \lambda s}  \int_{0}^{s} \mu_{u}^{D} du ds }{ t e^{ \lambda t  } } = \lim_{ t \to \infty }  \frac{ \int_{0}^{t} \mu_{s}^{D} ds }{ \lambda t + 1 }
=  \frac{ \overline{\mu}}{ \lambda},
\]
proving (\ref{x}).
\newline
\newline
\textit{Part II.} We claim that
\[
\lim_{t \to \infty} \frac{\int_{0}^{t} \left(  \delta_{I_K s} - \delta_{is} \right)  d W^{(1)}_s}{t} = 0.
\]
By definition (see (\ref{mgr_error})), it suffices to verify that
\[
\lim_{ t \to \infty  } \frac{ \int_{0}^{t}  \mu^{D}_{js} d W^{(1)}_s }{ t  } = 0
\]
holds for all $j=1,...,N.$ It is not hard to check by employing Lemma \ref{estimates} combined with the law of
large numbers, that the preceding limit does not change when the
functions $y_{iu}$, $ \frac{1}{y_{iu}} $ and $ \nu_{iu}$ are
substituted by $ e^{ ( \alpha_{i2} + \xi ) t }  $, $ e^{ - ( \alpha_{i2} +
\xi ) t }  $ and $ \alpha_{i2} ( \sigma^{D} )^2  $, respectively. In view of the latter observation, by definition (see (\ref{1})), we need to show that
\[
\lim_{t \to \infty} \frac{ \int_{0}^{t}
\left( \left( \overline{\mu}_i - \overline{\mu}  \right) \left( 1 - e^{ - \xi s } \right) +
\left( \mu_{0i} - \mu_0  \right) e^{ - \xi s } + \frac{ \xi \overline{\mu}_i }{ \xi + \alpha_{i2} }  \left(1 - e^{ - \left( \xi + \alpha_{i2}  \right) s}  \right) \right)  d W^{(1)}_s  }{ t }
\]
\[
+ \alpha_{i2} \lim_{t \to \infty}  \frac{ \int_{0}^{t} e^{ - \left( \xi + \alpha_{i2}  \right) s}
\int_{0}^{s} e^{ \left( \xi + \alpha_{i2}  \right) u} \left( \overline{\mu} + \left(
\mu_0 - \overline{\mu} \right) e^{- \xi u } \right) du  d W^{(1)}_s }{t }
\]
\[
+ \alpha_{i2} \sigma^{\mu}  \lim_{t \to \infty}  \frac{ \int_{0}^{t} e^{ - \left( \xi + \alpha_{i2} \right)s } \int_{0}^{s} e^{ \alpha_{i2}u } \int_{0}^{u} e^{ \xi x} d W^{(2)}_x du d W^{(1)}_s }{ t }
\]
\[
+ \sigma^{\mu} \phi_i
\lim_{t \to \infty}  \frac{ \int_{0}^{t} e^{ - \left(  \xi + \alpha_{i2} \right) s } \int_{0}^{s}
e^{ \left(  \xi + \alpha_{i2} \right) u } ds_u  dW^{(1)}_s }{t}  = 0.
\]
One checks that the first two terms vanish by the law of large numbers. The third and fourth limits vanish by part (iii) and (ii) of Lemma \ref{strassen}, respectively. This completes the proof of the second part.
\newline
\newline
\textit{Part III.} We have,
\[
\frac{1}{2} \lim_{t \to \infty} \frac{ \int_{0}^{t} \left( \delta^{2}_{is} - \delta^{2}_{I_K s} \right) ds }{ t } = \frac{1}{2}  \left(  \frac{\overline{\mu}_i - \overline{\mu}}{\sigma^{D}}  \right)^2
 +
 \frac{ \xi^2 + \left( \sigma^{\mu} / \sigma^{D} \right)^2 \left( 1 - \phi \phi_i \right)  }{ 2 \sqrt{ \xi^2 +
 \left( \sigma^{\mu} / \sigma^{D} \right)^2 \left( 1 - \phi^2_{i} \right)  }  }
\]
\[
- \frac{1}{2}  \left(  \frac{\overline{\mu}_{I_K} - \overline{\mu}}{\sigma^{D}}  \right)^2
 -
 \frac{ \xi^2 + \left( \sigma^{\mu} / \sigma^{D} \right)^2 \left( 1 - \phi \phi_{I_K} \right)  }{ 2 \sqrt{ \xi^2 +
 \left( \sigma^{\mu} / \sigma^{D} \right)^2 \left( 1 - \phi^2_{I_K} \right)  }  } .
\]
This can be derived by applying Lemmata \ref{strassen}, \ref{no_stationary}, \ref{stationary} and \ref{last}. The proof is now accomplished by combining the above three parts, some routine
algebraic transformations and the law of large numbers. $\qed$

\section{Interest Rate and Market Price of Risk: Further Long-Run Results}

The current section deals with asymptotic results for the interest rate and the market price of risk in heterogeneous economies. More precisely, it is shown that asymptotically, the latter parameters behave as those associated with a homogeneous economy populated by the dominating consumer. Under some mild conditions, we prove that the distance between these parameters in a heterogeneous economy and those associated with \textit{any} of the non-dominating consumer homogeneous economies, becomes unbounded as time goes to infinity.

\subsection{Market Price of Risk}
The next statement provides a full characterization of the market price of risk asymptotics in
heterogeneous economies.
\begin{Theorem}
(i) We have
\[
 \lim_{ t \to \infty } \left| \theta_t - \theta_{I_K t}  \right| = 0 .
\]
\newline
(ii) If $\phi_i = \phi_{I_K}$ for some $i \neq I_K$, then
\[
\lim_{t \to \infty} \left( \theta_t - \theta_{ i t}  \right) =
\sigma^{D} \left( \gamma_{I_K} - \gamma_i \right) - \frac{1}{\sigma^{D}} \left(
\overline{\mu}_{I_K} - \overline{\mu}_i \right).
\]
If $ \phi_i $ (for some $i \neq I_K$) is such that
\[
\frac{ \xi^2 + \left( \sigma^{\mu} / \sigma^{D} \right)^2 \left( 1 - \phi \phi_i \right)  }{ 2 \sqrt{ \xi^2 +
 \left( \sigma^{\mu} / \sigma^{D} \right)^2 \left( 1 - \phi^2_{i} \right)  }  }  \neq
 \frac{ \xi^2 + \left( \sigma^{\mu} / \sigma^{D} \right)^2 \left( 1 - \phi \phi_{I_K} \right)  }{ 2 \sqrt{ \xi^2 +
 \left( \sigma^{\mu} / \sigma^{D} \right)^2 \left( 1 - \phi^2_{I_K} \right)  }  } ,
 \]
then
\[
\limsup_{t \to \infty} \left| \theta_t - \theta_{ i t}  \right| =   \infty.
\]
\label{thm_bonds}
\end{Theorem}
\textbf{Proof.} (i) First, we shall prove
that $ \lim_{t \to \infty} \omega_{it} \theta_{it} = 0,$ for all
$i \neq I_K$. As in Section 5, the DDS theorem implies the existence
of a Brownian motion $ \left( \widetilde{B}(t) \right)_{t \in [ 0 , \infty)} $ such that
\[
e^{-at} \int_{0}^{t} e^{as} dB_s = e^{-at} \widetilde{B} \left(
\frac{ e^{ 2at} - 1 }{ 2a } \right),
\]
where $a>0$ is some constant and $\left( B_t \right)_{t \in [ 0 , \infty)} $ is a Brownian motion.
By exploiting the preceding fact, one checks
that $  \lim_{ t \to \infty } \frac{ \mu^{D}_{it} }{ \sqrt{ \log t } } < \infty, $
for all $i=0,...,N,$ which implies that
\begin{equation}
\limsup_{ t \to \infty
} \frac{ \theta_{jt}  }{ \sqrt{ \log t } } < \infty
\label{limmm}
\end{equation}
for all $ j=1,...,N.$ On the other hand, it was shown in
Theorem \ref{tolerance} that $ \omega_{it} \leq
 \frac{c_{it} }{ D_t } \max_{i=1,...,N} \gamma_{i} $ and all $ i \neq
I_K.$ We have in particular proved in Section 6 that $ \frac{c_{it}
}{ D_t } \leq e^{-a_i t }$, for some $a_i > 0 $, for all $i \neq I_K$.
This implies that
\begin{equation}
\omega_{it}  \leq  e^{- a_i t}  \max_{i=1,...,N} \gamma_{i}
\label{1'}
\end{equation}
holds for all $i \neq I_K$, and thus, by (\ref{limmm}), we have
$\omega_{it} \theta_{it} \leq  e^{- a'_i t}$
for all $ i \neq I_K$, and some constant $a'_i>0$. Therefore, by Proposition \ref{interest_mpr}, we have
\begin{equation}
\left| \theta_{t}  - \omega_{I_K t } \theta_{I_K t} \right| =
\sum_{i=1, i \neq I_K}^{N} \omega_{i t } \theta_{i t} \leq  \sum_{i=1, i \neq I_K}^{N} e^{- a'_i t} .
\label{2'}
\end{equation}
Finally, observe that $ \left| \theta_t - \theta_{ I_K t}  \right|
\leq \left| \theta_{t}  - \omega_{I_K t } \theta_{I_K t} \right| +
\sum_{i=1 , i \neq I_K}^{N} \omega_{ i t } \theta_{I_K t}, $ since
$\sum_{i=1}^{N} \omega_{it} = 1 $.
The proof of part (i) follows from (\ref{limmm}), (\ref{1'}) and (\ref{2'}).
\newline
(ii) If $\phi_i = \phi_{I_K},$ by part (i) we can
substitute $ \theta_T $ by $ \theta_{I_K T} .$ The assertion follows by noting that
\[
\lim_{ t \to \infty} \left| \theta_{I_K t } - \theta_{i t}  \right|  = \lim_{ t \to \infty} \left| \sigma^{D} \left( \gamma_i - \gamma_{I_K } \right) +
\frac{1}{ \sigma^{D} }  \left( \mu_{ I_K t } - \mu_{i t}  \right) \right|.
\]
Assume now that
\[
\frac{ \xi^2 + \left( \sigma^{\mu} / \sigma^{D} \right)^2 \left( 1 - \phi \phi_i \right)  }{ 2 \sqrt{ \xi^2 +
 \left( \sigma^{\mu} / \sigma^{D} \right)^2 \left( 1 - \phi^2_{i} \right)  }  }  \neq
 \frac{ \xi^2 + \left( \sigma^{\mu} / \sigma^{D} \right)^2 \left( 1 - \phi \phi_{I_K} \right)  }{ 2 \sqrt{ \xi^2 +
 \left( \sigma^{\mu} / \sigma^{D} \right)^2 \left( 1 - \phi^2_{I_K} \right)  }  }.
 \]
By part (i), the claim is equivalent to proving that
\begin{equation}
\limsup_{ t \to \infty } | \mu^{D}_{I_K t} - \mu^{D}_{i t } |  =  \infty.
\label{ccc}
\end{equation}
First, one checks by employing Lemma \ref{estimates} that the limit (\ref{ccc}) does not change
when substituting $\nu_{it}$, $y_{it}$ and $\frac{1}{ y_{it} }$ by $ \alpha_{i2} ( \sigma^{D})^2
$, $ \exp \left(
- \frac{ \alpha_{i2}  }{ \alpha_{i1}  } e^{ - \frac{ \alpha_{i2}  }{ \alpha_{i1}  }  }
\right) e^{ ( \alpha_{i2}  +  \xi) t } $ and $ \exp \left(
 \frac{ \alpha_{i2}  }{ \alpha_{i1}  } e^{ - \frac{ \alpha_{i2}  }{ \alpha_{i1}  }  }
\right) e^{ - ( \alpha_{i2}  +  \xi) t } $, respectively. Next, note that Fubini's theorem yields
\[
\frac{\alpha_{i2}}{ e^{  ( \xi + \alpha_{i2} ) T} }  \int_{0}^{T} e^{  \alpha_{i2}  u} \int_{0}^{u} e^{ \xi x } d W^{(2)}_x du  =
\]
\[
\frac{1}{e^{ \xi T } } \int_{0}^{T} e^{\xi u } dW^{(2)}_u   -  \frac{ 1 }{ e^{ ( \alpha_{i2} + \xi ) T  } } \int_{0}^{T} e^{ ( \alpha_{i2} + \xi) u }  dW^{(2)}_u .
\]
By exploiting the latter observations and the DDS theorem, one checks that
\[
\limsup_{t \to \infty} \left| \mu^{D}_{it} - \mu^{D}_{I_k  t} \right|  =
\limsup_{ t \to \infty} \left| f_i (t) - f_{I_k}  (t) \right|,
\]
where
\[
f_{i} (t) =
\frac{ 1 }{  \sqrt{ \left( \alpha_{i2} + \xi \right) t } }   \left(  \sigma^{D} \alpha_{i2} B^{i1} (t )  - \sigma^{ \mu } \left( \phi \phi_i - 1 \right) B^{i2} ( t )   + \sigma^{\mu} \phi_i
\sqrt{1 - \phi^2 } B^{i3}(t) \right) .
\]
Here, $B^{i1}(t), B^{i2}(t)$ and $B^{i3}(t)$ denote three independent Brownian motions.
By applying the DDS Theorem again, we can rewrite
\begin{equation}
f_{i} (t) = \frac{ 1 }{  \sqrt{ \left( \alpha_{i2} + \xi \right) t } } B^{(i)} \left( l_i t \right) ,
\label{f}
\end{equation}
where $B^{(i)}(t)$ is a Brownian motion, and
\[
l_i = \left( \sigma^{D} \alpha_{i2} \right)^2 + \left( \sigma^{\mu} \right)^2 \left(1- \phi \phi_i \right)^2
 + \left( \sigma^{\mu} \phi_i \right)^2 \left( 1 - \phi^2 \right) .
\]
Lastly, one checks that $\limsup_{ t \to \infty} \left| f_i (t) - f_{I_k}  (t) \right| = \infty$
by using the law of iterated logarithm and (\ref{f}), combined with the fact that
$ \frac{l_i}{ \alpha_{i2} + \xi } =
-2 \xi \sigma^{D}
+ 2 (\sigma^D)^2 \frac{ \xi^2 + \left( \sigma^{\mu} / \sigma^{D} \right)^2 \left( 1 - \phi \phi_i \right)  }{ 2 \sqrt{ \xi^2 +
 \left( \sigma^{\mu} / \sigma^{D} \right)^2 \left( 1 - \phi^2_{i} \right)  }  }
$. This completes the proof of Theorem \ref{thm_bonds}. $\qed$

\subsection{Interest Rate}

Analogously to Theorem \ref{thm_bonds}, we analyze in the next statement the asymptotics
of the interest rate in heterogeneous economies.

\begin{Theorem} (i) We have
\[
\lim_{t \to \infty } \left| r_t - r_{I_K t} \right| = 0.
\]
\newline
(ii) If $ \gamma_i = \gamma_{I_K}  $, $ \beta_i  = \beta_{I_K} $ and $ \phi_i  = \phi_{ I_K }$ for some $i \neq I_K$, then
\[
\lim_{ t \to \infty} \left( r_t - r_{ i t} \right) =
\rho_{I_K} - \rho_i + \gamma_{I_K} \left( \overline{ \mu }_{I_K}  - \overline{ \mu }_{i}  \right) .
\]
If at least one of the conditions: $ \gamma_i = \gamma_{I_K}  $, $ \beta_i  = \beta_{I_K} $ and
\[
\frac{ \xi^2 + \left( \sigma^{\mu} / \sigma^{D} \right)^2 \left( 1 - \phi \phi_i \right)  }{ 2 \sqrt{ \xi^2 +
 \left( \sigma^{\mu} / \sigma^{D} \right)^2 \left( 1 - \phi^2_{i} \right)  }  }  =
 \frac{ \xi^2 + \left( \sigma^{\mu} / \sigma^{D} \right)^2 \left( 1 - \phi \phi_{I_K} \right)  }{ 2 \sqrt{ \xi^2 +
 \left( \sigma^{\mu} / \sigma^{D} \right)^2 \left( 1 - \phi^2_{I_K} \right)  }  }
 \]
does not hold, for some $i \neq I_K$, then
\[
\limsup_{ t \to \infty} \left| r_t - r_{it} \right| = \infty.
\]
\label{thm_bonds2}
\end{Theorem}
\textbf{Proof.}
(i) By definition, we have
\[
r_t - \omega_{ i t} r_{ i t} = \sum_{ j = 1, j \neq I_K }^{N}
\omega_{jt} r_{jt} + \frac{1}{2} \sum_{ j=1}^{N} ( 1 - 1 / \gamma_j
) \omega_{jt} \left( \theta_{jt} - \theta_{t} \right)^2
\]
for all $i=1,...,N.$ We start by treating the second term.
Observe that Theorem \ref{tolerance}, part (i) of Theorem \ref{thm_bonds}, (\ref{ccc}) and
(\ref{1'}), imply that
\[
  \sum_{ j = 1 }^{N} ( 1 - 1 / \gamma_j
) \omega_{ jt } \left( \theta_{ jt } - \theta_{ t } \right)^2 \leq e^{ - a' t }
\]
for some constant $a'>0$. Next, note that (\ref{1'}) yields
\[
\sum_{ j=1, j \neq I_K }^{N} \left| \omega_{jt} r_{jt} \right|  \leq  \sum_{ j=1, j \neq  I_K }^{N} e^{ - a_j t } \left| r_{jt} \right|.
\]
As in the proof of Theorem \ref{surviving_consumer}, one can check that
$ \limsup_{t \to \infty} \frac{ r_{jt} }{t } < \infty $ for all $j=1,...,N$,
and thus we conclude that
\[
\left| r_{t} - \omega_{I_K t} r_{I_K t} \right|  \leq e^{ - a' t}
\]
for some constant $a'>0$. Finally, the proof of item (i) is accomplished by employing the inequality
$ | r_{t} - r_{ I_K t } | \leq | r_t - \omega_{ I_K t } r_{ I_K t}  | + r_{I_K t} | 1 -  \omega_{ I_K t } | $,
combined with the fact that $ 1 = \sum_{ j=1}^{N} \omega_{jt}$, (\ref{1'}) and the fact that
$ \limsup_{t \to \infty} \frac{ r_{jt} }{t } < \infty $ for all $j=1,...,N$.
\newline
(ii) If $\phi_i = \phi_{I_K}$, $ \gamma_i = \gamma_{I_K}$ and $ \beta_i = \beta_{I_K}$ for some $i \neq I_K$, the claim follows by combining part (i) with the fact that
\[
\left| r_{it} - r_{I_K t} \right|  =
\left| \rho_{I_K} - \rho_i
+ \gamma_{I_K} \left( \mu^{D}_{I_K t} - \mu^{D}_{it}  \right) \right|.
\]
Now, if at least one of the indicated conditions fails for some $i \neq I_K$,
the proof is in the same spirit as the one of item (ii) of Theorem \ref{thm_bonds}. The only distinction is as follows.
If $\lambda=\xi$, one can check that the problem can be reduced to proving that
\begin{equation}
\limsup_{t \to \infty } e^{ - \lambda t }  \left(
\sigma^{D} \int_{0}^{t} e^{ \lambda u } dW^{(1)}_u +
\int_{0}^{t} \int_{0}^{s} e^{ \lambda u} d W^{(2)}_u  ds  \right) = \infty .
\label{important}
\end{equation}
If $\lambda=0,$ we need to prove that
\[
\limsup_{t \to \infty }  \left(
\sigma^{D} W^{(1)}_t +
\int_{0}^{t} W^{(2)}_s  ds  \right) = \infty .
\]
Let $ G: C_{0}  \left( [0,1] ; \mathbb{R} \right) \to \mathbb{R}$ be a functional given by
$ G(f) = \int_{0}^{1} f(x) dx . $
Note that $G$ is continuous, since $| G(f) - G(g) | \leq || f - g ||_{\infty}$
holds for all $ f , g \in C_{0}  \left( [0,1] ; \mathbb{R} \right)$. By Strassen's functional law of iterated logarithm, we have
\[
\limsup_{N \to \infty } G \left(  \frac{1}{\sqrt{2 N \log \log N } } W^{(2)}_{Nx} \right)
= \limsup_{N \to \infty } \frac{ \int_{0}^{N} W^{(2)}_u du }{ N^{3 / 2} \sqrt{2 \log \log N} } = \max_{ f \in K^{(1)} } G ( f ) ,
\]
where the subspace $K^{(1)}$ is given in Definition \ref{cameron-martin}.
Note that $\max_{ f \in K^{(1)} } G ( f ) \geq G ( f_0 ) > 0,$ where $f_0(x)=x.$ The preceding observation combined with the fact
$ \lim_{t \to \infty} \frac{ W^{(1)}_t }{ t^{3 / 2} \sqrt{ \log \log t } } =0$ asserts that  $(\ref{important})$ holds
for $\lambda = 0$. Assume next that $\lambda \neq 0.$
By the DDS theorem, (\ref{important}) is equivalent to
\[
\limsup_{t \to \infty } e^{ - \lambda t }  \left(
\sigma^{D} B^{(1)} \left( \frac{e^{ 2 \lambda t} - 1 }{ 2 \lambda } \right)  +
\int_{0}^{t}
B^{(2)} \left(   \frac{ e^{ 2 \lambda s} - 1 }{ 2 \lambda } \right)  ds \right) = \infty,
\]
where $ B^{(1)} $ and $B^{(2)}$ denote two standard independent Brownian motions. By a change of variables, the claim is equivalent to
\begin{equation}
\limsup_{t \to \infty} \frac{1}{ \sqrt{t}} \left( \sigma^{D} B^{(1)}(t) + \int_{0}^{t} \frac{ B^{(2)} (u) }{1 + 2 \lambda u} du  \right)  = \infty .
\label{limm}
\end{equation}
The law of iterated logarithm yields
$
\lim_{t \to \infty} \frac{  \int_{1}^{t} \frac{ B^{(1)} (u) }{ u \left( 1 + 2 \lambda u \right) } du }{ \sqrt{t}} = 0 ,
$
and thus (\ref{limm}) can be rewritten as
\begin{equation}
\limsup_{t \to \infty} \frac{1}{ \sqrt{t}} \left( \sigma^{D} B^{(1)}(t) + \frac{1}{2 \lambda} \int_{1}^{t} \frac{ B^{(2)} (u) }{u} du  \right)  = \infty .
\label{verylast}
\end{equation}
Fix some $0 < \varepsilon < 1$. Consider the functional $H : C_{0} \left( [0,1] ; \mathbb{R}^2 \right) \to \mathbb{R},$ which is given by
\[
H ( f , g ) := \sigma^{D} f(1) + \frac{1}{2 \lambda} \int_{ \varepsilon}^{1} \frac{ g(u) }{ u } du.
\]
Note that $H$ is continuous, since $ \left| H( f , g ) - H( \widehat{f} , \widehat{g})  \right| \leq \sigma^{D} || f- \widehat{f} ||_{\infty} - \frac{\log \varepsilon }{ 2 \lambda }
|| g - \widehat{g} ||_{\infty}  $ is satisfied for all $ f , g , \widehat{f}, \widehat{g}
\in C_{0} \left( [0,1] ; \mathbb{R}^2 \right)  $. Next, Strassen's functional law of iterated logarithm yields
\[
\limsup_{ N \to \infty } H \left( \frac{1}{ \sqrt{2 N \log \log N } }  \left( B^{(1)}(N t) , B^{(2)}(N t) \right)  \right) = \max_{ \left( f , g  \right) \in K^{(2)} } H \left( f , g \right) ,
\]
where $K^{(2)}$ is introduced in Definition \ref{cameron-martin}. Observe that $\max_{ \left( f , g  \right) \in K^{(2)} } H \left( f , g \right) \geq H \left( h(x), h(x) \right) > 0$, where $h(x)=x.$ Therefore, we obtain that
\begin{equation}
 \limsup_{ N \to \infty }  \frac{1}{ \sqrt{2 N \log \log N } }  \left(  \sigma^{(D)} B^{(1)} \left( N \right) + \frac{1}{2 \lambda} \int_{ \varepsilon N}^{N} \frac{ B^{(2)} (u) }{ u } du  \right) > 0.
\label{last_lim}
\end{equation}
We claim next that
\[
\limsup_{ N \to \infty }  \frac{1}{ \sqrt{2 N \log \log N }}  \left(  \sigma^{(D)} B^{(1)} \left( N \right) + \frac{1}{2 \lambda} \int_{ 1 }^{N} \frac{ B^{ (2) } (u) }{ u } du  \right) > 0 .
\]
Assume towards contradiction that this is not the case. Then, Kolmogorov's 0-1 law implies that
\[
P \left( \limsup_{ N \to \infty }  \frac{1}{ \sqrt{2 N \log \log N }}  \left(  \sigma^{(D)} B^{(1)} \left( N \right) + \frac{1}{2 \lambda} \int_{ 1 }^{N} \frac{ B^{(2)} ( u ) }{ u } du  \right) > 0 \right) = 0 .
\]
Therefore, by exploiting the symmetry of the Brownian motion, we obtain that
\[
\lim_{ N \to \infty }  \frac{1}{ \sqrt{2 N \log \log N }}  \left(  \sigma^{(D)} B^{(1)} \left( N \right) + \frac{1}{2 \lambda} \int_{ 1 }^{N} \frac{ B^{(2)} (u) }{ u } du  \right) = 0,
\]
holds $P-$a.s. But, since $\sigma^{D}$ and $\lambda$ were arbitrary, we obtain that
\[
\limsup_{ N \to \infty }  \frac{1}{ \sqrt{2 N \log \log N }}  \left(  \sigma^{(D)} B^{(1)} \left( N \right) + \frac{1}{2 \lambda} \int_{ \varepsilon N  }^{N} \frac{ B^{(2)} (u) }{ u } du  \right) =
\]
\[
\limsup_{ N \to \infty }  \frac{1}{ \sqrt{2 N \log \log N }}
\bigg(
 \left( \sigma^{(D)}  - \sqrt{ \varepsilon }  \right) B^{(1)} \left( N \right) + \frac{1}{2 \lambda} \int_{ 1 }^{N} \frac{ B^{(2)} (u) }{ u } du
\]
\[
 + \widetilde{B}^{(1)} \left(  \varepsilon N \right)  - \frac{1}{2 \lambda} \int_{ 1 }^{ \varepsilon N} \frac{ B^{(2)} (u) }{ u } du
\bigg) = 0 ,
\]
where $\widetilde{B}^{(1)} \left(  t \right) = \sqrt{\varepsilon} B^{(1)} \left( \frac{t}{\varepsilon} \right) $ is a Brownian motion (independent of $B^{(2)}$), and $\varepsilon>0$ is sufficiently small. This is a contradiction to (\ref{last_lim}) proving (\ref{limm}). $\qed$
\newline
\newline
\textbf{Acknowledgments.} It is my pleasure to thank my supervisor Semyon Malamud for introducing me to the topic of 'natural selection in financial markets'.
I am indebted to him for numerous fruitful conversations and for detailed remarks on the preliminary version of the manuscript.
I would also like to thank Yan Dolinsky, Mikhail Lifshits, Johannes Muhle-Karbe and Chris Rogers for useful discussions,
and Kerry Back and two anonymous referees for helpful remarks that lead to a substantial improvement of the paper.


\begin{thebibliography}{Bow75}


\itemsep=\smallskipamount

\bibitem{Abel} {\sc Abel, A.B.:}
Asset prices under habit formation and catching up with the Joneses.
{\em American Econ. Rev.} {\bf 80}(2), 38-42, (1990)

\bibitem{Back} {\sc Back, K.:}
Incomplete and asymmetric information in asset pricing theory.
{\em Stochastic Methods in Finance}, Lecture Notes in Mathematics, Springer, 1-23 (2004)

\bibitem{Basak1} {\sc Basak, S.:}
A model of dynamic equilibrium asset pricing with
heterogeneous beliefs and extraneous risk.
{\em  J. Econ. Dyn. Control} {\bf 24}, 63-95 (2000)

\bibitem{Basak2} {\sc Basak, S.:}
Asset pricing with heterogeneous beliefs.
{\em  J. Bank. Finance} {\bf 29}, 2849-2881 (2005)


\bibitem{BU}
{\sc Bhamra, H.S., R. Uppal:}
The effect of introducing a non-redundant derivative on the
volatility of stock-market returns when agents differ in risk aversion.
{\em Rev. Financ. Stud.} {\bf 22}, 2303-2330 (2009)

\bibitem{BU2}
{\sc Bhamra, H.S., R. Uppal:}
Asset prices with heterogeneity in preferences and beliefs.
{\em Working paper} (2010)


\bibitem{BE} {\sc Blume, L., Easley, D.:}
If you are so smart, why aren’t you rich.
Belief selection in complete and incomplete markets.
{\em  Econometrica} {\bf 74}, 929-966 (2006)

\bibitem{CJMN} {\sc  Cvitani\'c, J., Jouini, E., Malamud, S., Napp, C.:}
Financial markets equilibrium with heterogeneous agents.
{\em To appear in Rev. Finance}


\bibitem{CK} {\sc  Chan, Y.L., Kogan, L.:}
Catching up with the Joneses: Heterogeneous preferences and
the dynamics of asset prices.
{\em J. Pol. Econm.} {\bf 110}, 1255-1285 (2002)


\bibitem{CM} {\sc  Cvitani\'c, J., Malamud, S.:}
Relative extinction of heterogeneous agents.
{\em B. E. J. Theor. Econom.}  {\bf 10}, article 4 (2010)

\bibitem{CM2} {\sc  Cvitani\'c, J., Malamud, S.:}
Price impact and portfolio impact.
{\em J. Finan. Econom.}  {\bf 100}, 201-225 (2011)

\bibitem{DH} {\sc Duffie, D., Huang, C.-F.:} Implementing Arrow-Debreu equilibria by
continuous trading of few long-lived securities. {\it Econometrica} {\bf 53},
1337-1356 (1985)

\bibitem{D} {\sc Dumas, B.:}
Two person dynamic equilibrium in the capital market.
{\em J. Rev. Financ. Stud.} {\bf 2}, 157-188 (1989)


\bibitem{DKU} {\sc Dumas, B., Kurshev, A., Uppal, R.:}
Equilibrium portfolio strategies in the
presence of sentiment risk and excess volatility.
{\em J. Finance} {\bf 64}, 579-629 (2009)

\bibitem{FHLW} {\sc Fedyk, Y., Heyerdahl-Larsen, C., Walden, J.:}
Market selection and welfare in multi-asset economies.
{\it Unpublished working paper,} University of California at Berkeley (2010)


\bibitem{Fr}{\sc Friedman, B. M.:}
Money, Credit, and Interest Rates in The Business Cycle.
{\it In: R.J. Gordon, Editor, The American Business Cycle: Continuity and Change, National Bureau of Economic Research Studies in Business Cycles} {\bf 25}, University of Chicago Press, Chicago,
395-458 (1986)



\bibitem{HMT} {\sc Hugonnier, E., Malamud. S., Trubowitz, E.:}
Endogenous completeness of diffusion driven equilibrium markets.
{\em Working paper} (2011)


\bibitem{JMN} {\sc Jouini, E., Marin, J.-M., Napp, C.:}
Discounting and divergence of opinion.
{\em J. Econom. Theory} {\bf 145}, 830-859 (2010)


\bibitem{JN1} {\sc Jouini, E., Napp, C.:}
Consensus consumer and intertemporal
asset pricing under heterogeneous beliefs.
{\em Rev. Econm. Stud.} {\bf 74}, 1149-1174 (2007)


\bibitem{JN2} {\sc Jouini, E., Napp, C.:}
Unbiased disagreement and the efficient
market hypothesis. {\em To appear in Rev. Finance}


\bibitem{KRWW1} {\sc
Kogan, L., Ross, S., Wang, J., Westerfield, M.:}
The price impact and survival of irrational traders.
{\em J. Finance} {\bf 61}, 195-229 (2006)


\bibitem{KRWW2} {\sc
Kogan, L., Ross, S., Wang, J., Westerfield, M.:}
Market Selection. {\em Working Paper} (2011)


\bibitem{LS1} {\sc Liptser, R.S., Shiryaev, A.N.:}
Statistics of Random Processes: General Theory, 2nd edn.
{\em Springer} (2001)


\bibitem{LS2} {\sc Liptser, R.S., Shiryaev, A.N.:}
Statistics of Random Processes: Applications, 2nd edn.
{\em Springer} (2001)


\bibitem{NR} {\sc Nishide, K., Rogers, L.C.G.:}
Market selection: hungry misers and happy bankrupts.
{\em To appear in Mat. Financ. Econom.}


\bibitem{RY} {\sc
Revuz, D., Yor, M.:}
Continuous Martingales and Brownian Motion, 3rd edn.
{\em Springer} (1999)


\bibitem{Sa} {\sc Sandroni, A.:}
Do markets favor agents able to make accurate
predictions?
{\em Econometrica} {\bf 68}, 1303-1334 (2000)


\bibitem{SX} {\sc Scheinkman, J., Xiong, W.:}
Over confidence and Speculative Bubbles.
{\em  J. Pol. Econ. } {\bf 111}, 1183-1219 (2003)

\bibitem{W} {\sc Wang, J.:}
The term structure of interest rates in a pure exchange economy with heterogeneous investors.
{\em  J Finan. Econm.} {\bf 41}, 75-100 (1996)


\bibitem{XY} {\sc Xiong, W., Yan, H.:}
Heterogeneous expectations and bond
markets. {\em  Rev. of Finan. Stud.} {\bf 23}, 1405-1432 (2010)


\bibitem{XZ} {\sc Xiouros, C., Zapatero, F.:}
The representative agent of an economy with external habit-formation and heterogeneous risk-aversion {\em  Rev. of Finan. Stud.} {\bf 23}, 3017-3047 (2010)


\bibitem{Y1} {\sc Yan, H.:}
Natural selection in financial markets: Does it work?
{\em Management Science} {\bf 54}, 1935-1950 (2008)


\bibitem{Y2} {\sc Yan, H.:}
Is noise trading canceled out by aggregation?
{\em  Management Science} {\bf 56}, 1047--1059 (2010)

\end{thebibliography}
\end{document}